\documentclass[pageno]{vldb-arxiv-v2}
\usepackage{graphicx,mathtools,kantlipsum}
\usepackage[normalem]{ulem}
\usepackage{url}
\usepackage{amsmath,amssymb,amsfonts}
\usepackage{listings}
\usepackage{color}
\usepackage{courier}
\usepackage{float}
\usepackage{multirow}
\usepackage{makecell}
\usepackage{algorithm} 
\usepackage[noend]{algpseudocode}
\usepackage{todonotes}
\usepackage{booktabs}
\usepackage{caption}
\usepackage{subcaption}
\usepackage{hyperref}
\usepackage[capitalize]{cleveref}
\usepackage{balance}
\usepackage{enumitem}
\newcommand{\hlc}[1]{\textcolor{black}{#1}}

\vldbTitle{Pangolin: An Efficient and Flexible Parallel Graph Pattern Mining System on CPU and GPU}
\vldbAuthors{Xuhao Chen, Roshan Dathathri, Gurbinder Gill, Keshav Pingali}
\vldbDOI{https://doi.org/10.14778/xxxxxxx.xxxxxxx}
\vldbVolume{1}
\vldbNumber{xxx}
\vldbYear{2020}

\begin{document}

\title{Pangolin: An Efficient and Flexible Graph Pattern Mining System on CPU and GPU \vspace{-10pt}}
\numberofauthors{1}
\author{
	\alignauthor Xuhao Chen, Roshan Dathathri, Gurbinder Gill, Keshav Pingali\\
	\affaddr{The University of Texas at Austin}\\
	\email{cxh@utexas.edu, \{roshan,gill,pingali\}@cs.utexas.edu}
}

\date{}
\maketitle
\begin{abstract}
There is growing interest in \hlc{graph pattern mining (GPM)} problems such as motif counting.
\hlc{GPM} systems have been developed to provide unified interfaces
for programming algorithms for these problems and for running them on parallel
systems. However, existing systems may take hours to mine even simple patterns 
in moderate-sized graphs, which significantly limits their real-world usability.

We present \emph{Pangolin}, a high-performance and flexible in-memory
GPM framework targeting shared-memory CPUs and GPUs.
Pangolin is the \hlc{first GPM system that provides high-level abstractions for GPU processing}.
It provides a simple programming interface based on the
extend-reduce-filter model, which enables users to specify
application-specific knowledge for search space pruning
and isomorphism test elimination. We describe novel optimizations 
that exploit locality, reduce memory consumption, and mitigate 
the overheads of dynamic memory allocation and synchronization.

Evaluation on a 28-core CPU demonstrates that
Pangolin outperforms
\hlc{existing GPM frameworks Arabesque, RStream, and Fractal
by 49$\times$, 88$\times$, and 80$\times$ on average, respectively.}
Acceleration on a V100 GPU further improves performance
of Pangolin by 15$\times$ on average.
Compared to state-of-the-art \hlc{hand-optimized GPM} applications,
Pangolin provides competitive performance with less programming effort.
\end{abstract}

\section{Introduction}\label{sect:intro}
\begin{figure*}[t]
\centering
	\begin{minipage}[t]{0.33\linewidth}
		\includegraphics[width=\textwidth]{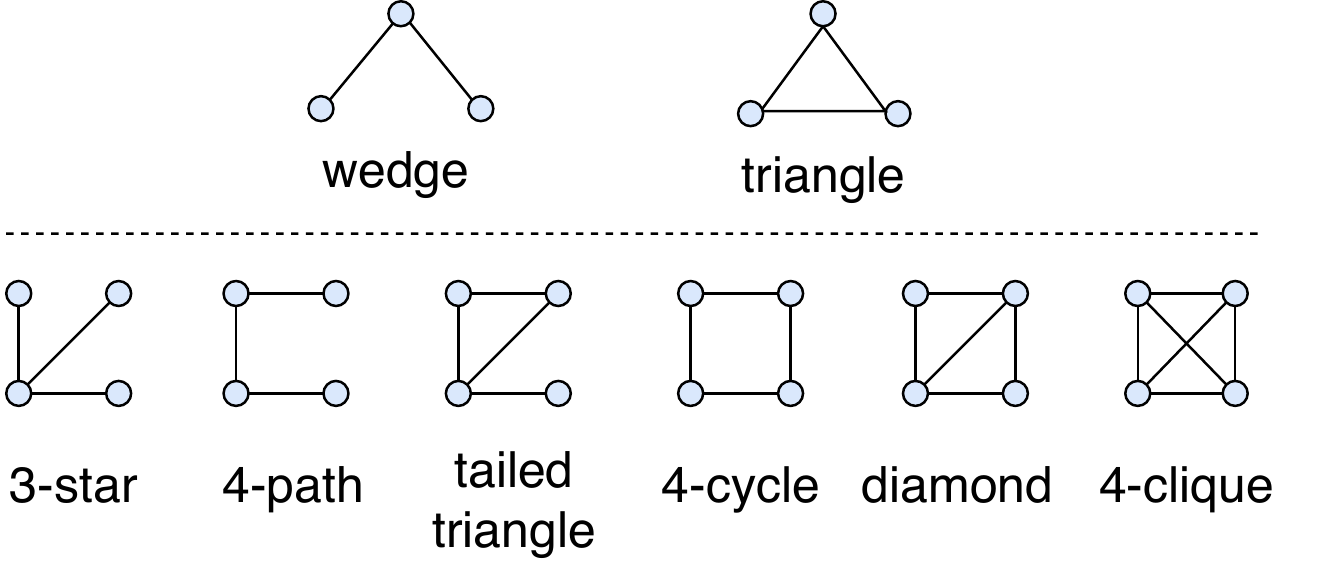}
		\vspace{-0.37cm}
		\caption{\small 3-vertex motifs (top) and 4-vertex motifs (bottom).}
		\vspace{-0.5cm}
		\label{fig:4-motifs}
	\end{minipage}
\hfill
	\begin{minipage}[t]{0.37\linewidth}
		\includegraphics[width=\textwidth]{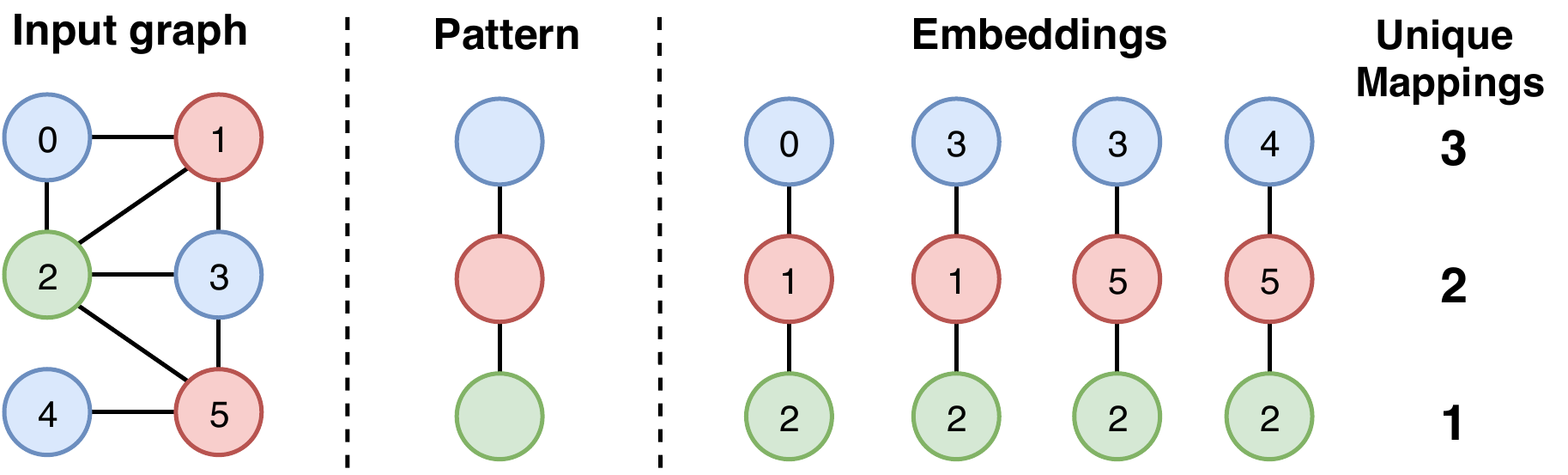}
		\vspace{-0.37cm}
		\caption{\small An example of the GPM problem.}
		\vspace{-0.5cm}
		\label{fig:example}
	\end{minipage}
\hfill
	\begin{minipage}[t]{0.23\linewidth}
		\includegraphics[width=\textwidth]{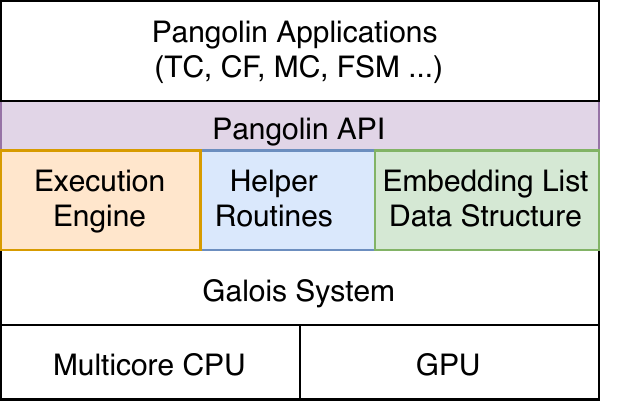}
		\vspace{-0.37cm}
		\caption{\small System overview of Pangolin (shaded parts).}
		\vspace{-0.5cm}
		\label{fig:overview}
	\end{minipage}
	\vspace{-0.1cm}
\end{figure*}

Applications that use graph data are becoming increasingly important in 
many fields such as world wide web, advertising, social media, and biology.
Graph analytics algorithms such as PageRank and SSSP have been studied
extensively and many frameworks have been proposed to provide both high
performance and high productivity~\cite{Pregel,GraphLab,Galois,Ligra}.
Another important class of graph problems deals with \hlc{\emph{graph pattern mining} (GPM)},
which has plenty of applications in areas such as chemical engineering~\cite{chemical},
bioinformatics~\cite{Motifs1,Protein}, and social sciences~\cite{Social}.
GPM discovers relevant patterns in a given graph.
One example is \emph{triangle counting}, 
which is used to mine graphs in security applications~\cite{Voegele2017}. 
Another example is \emph{motif counting}~\cite{Motifs2,Motif3}, 
which counts the frequency of certain structural patterns; 
this is useful in evaluating network models or classifying vertex roles. 
\cref{fig:4-motifs} illustrates the 3-vertex and 4-vertex motifs.

Compared to graph analytics, GPM algorithms
are more difficult to implement on parallel platforms; 
for example, unlike graph analytics algorithms, 
they usually generate enormous amounts of intermediate data. 
GPM systems such as \hlc{Arabesque~\cite{Arabesque}, RStream~\cite{RStream}, and Fractal~\cite{Fractal}}
have been developed to provide abstractions for improving programmer productivity.
Instead of the vertex-centric model used in graph analytics systems~\cite{Pregel},
Arabesque proposed an \textit{embedding-centric} programming model.
In Arabesque, computation is applied on individual
embeddings (\emph{i.e.}, subgraphs) concurrently.
It provides a simple programming interface that substantially
reduces the complexity of application development.
However, existing systems suffer dramatic performance 
loss compared to hand-optimized implementations.
For example, Arabesque and RStream take 98s and 39s respectively to count
3-cliques for the \texttt{Patent} graph with 2.7M vertices and 28M edges,
while a custom solver (KClist)~\cite{KClique} counts it in 0.16s.
This huge performance gap significantly limits the usability of
existing GPM frameworks in real-world applications.

The first reason for this poor performance is that existing 
GPM systems provide limited support for application-specific
customization. The state-of-the-art systems focus on generality and
provide high-level abstraction to the user for ease-of-programming.
Therefore, they hide as many execution details as possible from the user, 
which substantially limits the flexibility for algorithmic customization.
The complexity of GPM algorithms is primarily due 
to combinatorial enumeration of embeddings and
isomorphism tests to find canonical patterns.
Hand-optimizing implementations exploit application-specific
knowledge to aggressively prune the enumeration 
search space or elide isomorphism tests or both.
Mining frameworks need to support such optimizations
to match performance of hand-optimized applications.

The second reason for poor performance is inefficient implementation of
parallel operations and data structures. Programming parallel processors
requires exploring trade-offs between synchronization overhead, memory management,
load balancing, and data locality. However, the state-of-the-art
GPM systems target either distributed or out-of-core platforms,
and thus are not well optimized for shared-memory multicore/manycore architectures.


In this paper, we present Pangolin, an efficient in-memory GPM framework.
Pangolin provides a simple yet flexible
embedding-centric programming interface,
based on the {\it extend-reduce-filter} model,
which enables application-specific customization (Section~\ref{sect:design}).
Application developers can implement aggressive pruning
strategies to reduce the enumeration search space, and
apply customized pattern classification methods to elide
generic isomorphism tests (Section~\ref{sect:app-opt}).

To make full use of parallel hardware, we optimize parallel
operations and data structures, and provide helper routines
to the users to compose higher level operations.
Pangolin is built as a lightweight layer on top of
the Galois~\cite{Galois} parallel library and
LonestarGPU~\cite{LonestarGPU} infrastructure,
targeting both shared-memory multicore CPUs and GPUs.
Pangolin includes novel optimizations that exploit locality,
reduce memory consumption, and mitigate overheads of
dynamic memory allocation and synchronization (Section~\ref{sect:impl}).

Experimental results (Section~\ref{sect:eval}) on a 28-core CPU demonstrate 
that Pangolin outperforms existing \hlc{GPM frameworks, 
Arabesque, RStream, and Fractal, by 49$\times$, 88$\times$, and 80$\times$ on average, respectively.}
Furthermore, Pangolin on V100 GPU outperforms
Pangolin on 28-core CPU by 15$\times$ on average.
Pangolin provides performance competitive to
state-of-the-art hand-optimized \hlc{GPM} applications,
but with much less programming effort.
To mine 4-cliques in a real-world web-crawl graph (gsh)
with 988 million vertices and 51 billion vertices,
Pangolin takes $\sim6.5$ hours on a 48-core Intel Optane PMM
machine~\cite{optane} with 6 TB (byte-addressable) memory.
To the best of our knowledge, this is the largest graph
on which 4-cliques have been mined.

In summary, Pangolin makes the following contributions:

\vspace{-0.1cm}
\begin{itemize}[leftmargin=*]
\setlength\itemsep{-0.2em}

\item We investigate the performance gap between state-of-the-art 
\hlc{GPM systems} and hand-optimized approaches, and point 
out two key features absent in existing systems: 
\emph{pruning enumeration space} and \emph{eliding isomorphism tests}.

\item We present a high-performance in-memory GPM system,
Pangolin, which enables \emph{application-specific optimizations} and
provides transparent parallelism on CPU or GPU.
To the best of our knowledge, it is \hlc{the first 
GPM system that provides high-level abstractions 
for GPU processing}.

\item We propose novel techniques that enable the user to aggressively 
prune the enumeration search space and elide isomorphism tests.




\item We propose novel optimizations that exploit locality, 
reduce memory usage, and mitigate overheads of dynamic 
memory allocation and synchronization on CPU and GPU.


\item We evaluate Pangolin on a multicore CPU and a GPU
to demonstrate that Pangolin is substantially faster than 
existing GPM frameworks. \hlc{Compared to hand-optimized applications,
it provides competitive performance while requiring less programming effort.}

\end{itemize}

\section{Background and Motivation}\label{sect:back}
We describe GPM concepts, applications,
as well as algorithmic and architectural optimizations 
in state-of-the-art hand-optimized GPM solvers.
Lastly, we point out performance limitations of existing GPM frameworks.

\subsection{Graph Pattern Mining}\label{sect:gpm}
\hlc{Given an \emph{input graph} $G$ and a \emph{pattern} $P$ 
which is a subgraph defined by the user (e.g., triangle or clique),
the goal of GPM is to find the \emph{embeddings}, i.e.,
subgraphs in $G$ which are isomorphic to $P$.}
In the input graph in \cref{fig:example}, 
colors represent vertex labels, and numbers denote vertex IDs.
The 3-vertex pattern is a blue-red-green chain, 
and there are four embeddings of this pattern in the input graph, 
shown on the right of the figure. 
In a specific GPM problem, \hlc{the user may be interested in
some statistical information (i.e., pattern frequency), 
instead of listing all the embeddings.}
The measure of the frequency of $P$ in $G$, 
termed \textit{support}, is also defined by the user.
For example, in triangle counting, 
the support is defined as the total count of triangles.
\hlc{In some problems, the user might be interested in multiple patterns.
In this work, we focus on connected patterns only.}

There are two types of GPM problems targeting two types of embeddings. 
In a \textit{vertex-induced} embedding, 
a set of vertices is given and the subgraph of interest is obtained from these 
vertices and the set of edges in the input graph connecting these vertices. 
Triangle counting uses vertex-induced embeddings. In an \textit{edge-induced} embedding, 
a set of edges is given and the subgraph is formed by including all the endpoints of these 
edges in the input graph. Frequent subgraph mining (FSM) is an edge-induced GPM problem.

A GPM algorithm enumerates embeddings of the given pattern(s). 
If duplicate embeddings exist ({\it automorphism}), 
the algorithm chooses one of them as the {\it canonical} one
(namely canonical test) and collects statistical information 
about these canonical embeddings such as the total count. 
The canonical test needs to be performed on each embedding, 
and can be complicated and expensive for complex problems such as FSM.
Enumeration of embeddings in a graph grows exponentially 
with the embedding size (number of vertices or edges in the embedding), 
which is computationally expensive and consumes lots of memory. 
In addition, a graph isomorphism (GI) test is needed for each embedding 
to determine whether it is \textit{isomorphic} to a pattern. 
Unfortunately, the GI problem is not solvable in polynomial time~\cite{Garey}.
It leads to compute and memory intensive algorithms~\cite{Bliss}
that are time-consuming to implement.

\hlc{Graph analytics problems typically involve 
allocating and computing 
labels on vertices or edges of the input graph iteratively. 
On the other hand, GPM problems involve generating 
embeddings of the input graph 
and analyzing them. 
Consequently, GPM problems require much more memory 
and computation to solve.
The memory consumption is not only proportional to the graph size, 
but also increases exponentially as the embedding size increases~\cite{Arabesque}.
Furthermore, GPM problems require compute-intensive operations, 
such as isomorphism test and automorphism test on each embedding.
Thus, GPM algorithms are more difficult to develop, 
and conventional graph analytics systems 
\hlc{~\cite{GRAPE,EvoGraph,SIMD-X,Khorasani,MultiGraph,Falcon,PowerGraph,Gemini,Gluon}}
are not sufficient to provide a good trade-off between programmability and efficiency.}

\subsection{Hand-Optimized GPM Applications}\label{sect:handopt}

We consider 4 applications: triangle counting (TC), 
clique finding (CF), motif counting (MC), and frequent subgraph mining (FSM). 
Given the input graph which is undirected,
TC counts the number of triangles while CF enumerates all complete subgraphs
\footnote{\scriptsize A $k$-vertex complete subgraph is a connected subgraph in which 
each vertex has degree of $k-1$ (\emph{i.e.}, any two vertices are connected).}
(i.e., cliques) contained in the graph. 
TC is a special case of CF as it counts 3-cliques.
MC counts the number of occurrences (i.e., frequency) of each structural 
pattern (also known as \emph{motif} or \emph{graphlet}). 
As listed in \cref{fig:4-motifs}, $k$-clique is one of the patterns in $k$-motifs. 

FSM finds frequent patterns in a \emph{labeled} graph. 
A measure of frequency called {\em support} is provided 
by the application developer, 
and all patterns with support above a given threshold are 
considered to be frequent and must be discovered. 
A simple definition of support is the count of the 
embeddings associated with the pattern (used in TC, CF, MC).
A more widely used support definition is \textit{minimum image-based} 
(MNI) support (a.k.a. \emph{domain support}),
which has the anti-monotonic property 
\footnote{\scriptsize The support of a supergraph should not 
exceed the support of a subgraph; this allows the GPM algorithm to stop extending 
embeddings as soon as they are recognized as infrequent.}.
It is calculated as the minimum number of distinct mappings for any 
vertex (i.e., domain) in the pattern over all embeddings of the pattern. 
In \cref{fig:example}, the MNI support of the pattern is $min\{3,2,1\} = 1$.

Several hand-optimized implementations exist for each of these 
applications on multicore CPU~\cite{Shun,PGD,BeyondTri,Bressan,ParFSM},
GPU~\cite{Green,MotifGPU,CUDA-MEME,GpuFSM}, distributed 
CPU~\cite{Suri,PDTL,DistGraph}, and multi-GPU~\cite{TriCore,DistTC,Rossi}.
They employ application-specific optimizations to 
reduce algorithm complexity.
The complexity of GPM algorithms is primarily due to two aspects: 
combinatorial enumeration and isomorphism test.
Therefore, hand-optimized implementations focus on either \emph{pruning
the enumeration search space} or \emph{eliding isomorphism test} or both. 
We describe some of these techniques briefly below.

\textbf{Pruning Enumeration Search Space:}
In general GPM applications, 
\hlc{new embeddings are generated by extending existing embeddings 
and then they may be discarded because they are either 
not interesting or a duplicate ({\it automorphism}).}
However, in some applications like CF~\cite{KClique}, 
duplicate embeddings can be detected 
eagerly before extending current embeddings, 
based on properties of the current embeddings. 
We term this optimization as {\it eager pruning}.
Eager pruning can significantly reduce the search space.
Furthermore, the input graphs are converted into directed acyclic
graphs (DAGs) in state-of-the-art 
TC~\cite{TriCore}, CF~\cite{KClique}, and MC~\cite{ESCAPE} 
solvers, to significantly reduce the search space.

\textbf{Eliding Isomorphism Test:}
In most hand-optimized TC, CF, and MC solvers,
isomorphism test is completely avoided by taking advantage of the
pattern characteristics. For example, a parallel MC solver, PGD~\cite{PGD},
uses an ad-hoc method for a specific $k$. Since it only counts 3-vertex 
and 4-vertex motifs, all the patterns (two 3-motifs and six 4-motifs 
as shown in \cref{fig:4-motifs}) are known in advance. 
Therefore, some special (and thus easy-to-count) patterns (\emph{e.g.,} 
cliques\footnote{\scriptsize Cliques can be identified by checking connectivity among 
vertices without generic isomorphism test.}) are counted first, and the frequencies 
of other patterns are obtained in constant time using the relationship 
among patterns\footnote{\scriptsize For example, the count of diamonds can be 
computed directly from the counts of triangles and 4-cliques~\cite{PGD}.}. 
In this case, no isomorphism test is needed, 
which is typically an order-of-magnitude faster~\cite{PGD}. 


\textbf{Summary:}
Most of the algorithmic optimizations exploit application-specific 
knowledge, which can only be enabled by application developers. 
A generic GPM framework should be flexible enough to allow users 
to compose as many of these optimization techniques as possible,
and provide parallelization support for ease of programming. 
Pangolin is the first GPM framework to do so.

\subsection{Existing GPM Frameworks}

Existing GPM systems target 
either distributed-memory~\cite{Arabesque,Fractal,ASAP} or 
out-of-core~\cite{RStream,Kaleido,AutoMine} platforms, 
and they make tradeoffs specific for their targeted architectures. 
None of them target in-memory GPM on a multicore CPU 
or a GPU. 
Consequently, they do not pay much attention to
reducing the synchronization overheads among threads 
within a CPU/GPU or reducing memory consumption overheads.
Due to this, naively porting these GPM systems to 
run on a multicore CPU or GPU would 
lead to inefficient implementations. 
We first describe two of these GPM systems briefly 
and then discuss their major limitations.


Arabesque~\cite{Arabesque} is a distributed GPM system. 
It proposes ``think like an embedding'' (TLE) programming paradigm, 
where computation is performed in an embedding-centric manner. 
It defines a \textit{filter-process} computation model which 
consists of two functions: (1) \textit{filter}, which indicates 
whether an embedding should be processed and (2) \textit{process}, 
which examines an embedding and may produce some output. 

RStream~\cite{RStream} is an out-of-core single-machine GPM system. 
Its programming model is based on relational algebra. 
Users specify how to generate embeddings using relational 
operations such as \texttt{\small select}, \texttt{\small join}, and \texttt{\small aggregate}. 
It stores intermediate data (\emph{i.e.}, embeddings) 
on disk while the input graph is kept in memory for reuse.
It streams data (or table) from disk and 
uses relational operations that may produce more 
intermediate data, which is stored back on disk.

\textbf{Limitations in API:}
{\emph{Most of the application-specific optimizations 
like pruning enumeration search space and avoiding isomorphism
test are missing in existing GPM frameworks}}, as they focus 
on providing high-level abstractions but lack support for 
application-specific customization.
The absence of such key optimizations in existing systems 
results in a huge 
performance gap when compared to hand-optimized implementations.
Moreover, some frameworks like Rstream support only 
edge-induced embeddings 
but for applications like CF, 
the enumeration search space is much smaller 
using vertex-induced exploration than edge-induced one. 

\textbf{Data Structures for Embeddings:}
Data structures used to store embeddings in existing GPM systems are not efficient.
Both Arabesque and RStream store embeddings in an array of structures (AoS), 
where the embedding structures consists of a vertex set and an edge set.
Arabesque also proposes a space efficient data structure called the \emph{Overapproximating
Directed Acyclic Graph} (ODAG), but it requires extra canonical test for each embedding, 
which has been demonstrated to be very expensive for large graphs~\cite{Arabesque}.

\textbf{Materialization of Data Structures:} 
The list or array of intermediate embeddings in both Arabesque 
and RStream is always materialized in memory and in disk, respectively. 
This has significant overheads as the size of such data 
grows exponentially. Such materialization may not be needed 
if the embeddings can be filtered or processed immediately. 

\textbf{Dynamic Memory Allocation:} 
As the number of (intermediate) embeddings are not known before 
executing the algorithm, memory needs to be 
allocated dynamically for them. Moreover, during parallel execution, 
different threads might allocate memory for embeddings they create or enumerate. 
Existing systems use standard ({\tt std}) maps and sets, 
which internally use a global lock to dynamically allocate memory.
This limits the performance and scalability.

\textbf{Summary:}
Existing GPM systems have limitations in their API, 
execution model, and implementation. 
Pangolin addresses these issues by 
permitting application-specific optimizations in its API, 
optimizing the execution model, and 
providing an efficient, scalable implementation 
on multicore CPU and GPU.
These optimizations can be applied to existing 
embedding-centric systems like Arabesque.

\definecolor{codegreen}{rgb}{0,0.6,0}
\definecolor{codegray}{rgb}{0.5,0.5,0.5}
\definecolor{codepurple}{rgb}{0.58,0,0.82}
\definecolor{backcolour}{rgb}{1,1,1}
\lstdefinestyle{mystyle}{
    backgroundcolor=\color{backcolour},
    commentstyle=\color{codegreen},
    keywordstyle=\color{blue},
    numberstyle=\tiny\color{codegray},
    stringstyle=\color{codepurple},
    basicstyle=\scriptsize\ttfamily,
    breakatwhitespace=false,
    breaklines=true,
    captionpos=b,
    keepspaces=true,
    numbers=left,
    numbersep=5pt,
    showspaces=false,
    showstringspaces=false,
    showtabs=false,
    tabsize=2
}
\lstset{style=mystyle}
\lstset{aboveskip=0pt,belowskip=-5pt}

\section{Design of Pangolin Framework}\label{sect:design}
\hlc{\cref{fig:overview} illustrates an overview of the Pangolin system.
Pangolin provide a simple API (purple box) to the user for writing GPM applications.
The unified execution engine (orange box) follows the embedding-centric model.
Important common operations are encapsulated 
and provided to the user 
in the helper routines (blue box), 
which are optimized for both CPU and GPU.
The embedding list data structure (green box) is also optimized for
different architectures to exploit hardware features. 
Thus,
Pangolin hides most of the architecture oriented programming complexity
and achieves high performance and high productivity simultaneously.
In this section, we describe the execution model, programming interface (i.e., API), 
and example applications of Pangolin.}

\begin{figure*}[t]
\centering
	\begin{minipage}[t]{0.32\linewidth}
	\centering
	\includegraphics[width=\textwidth]{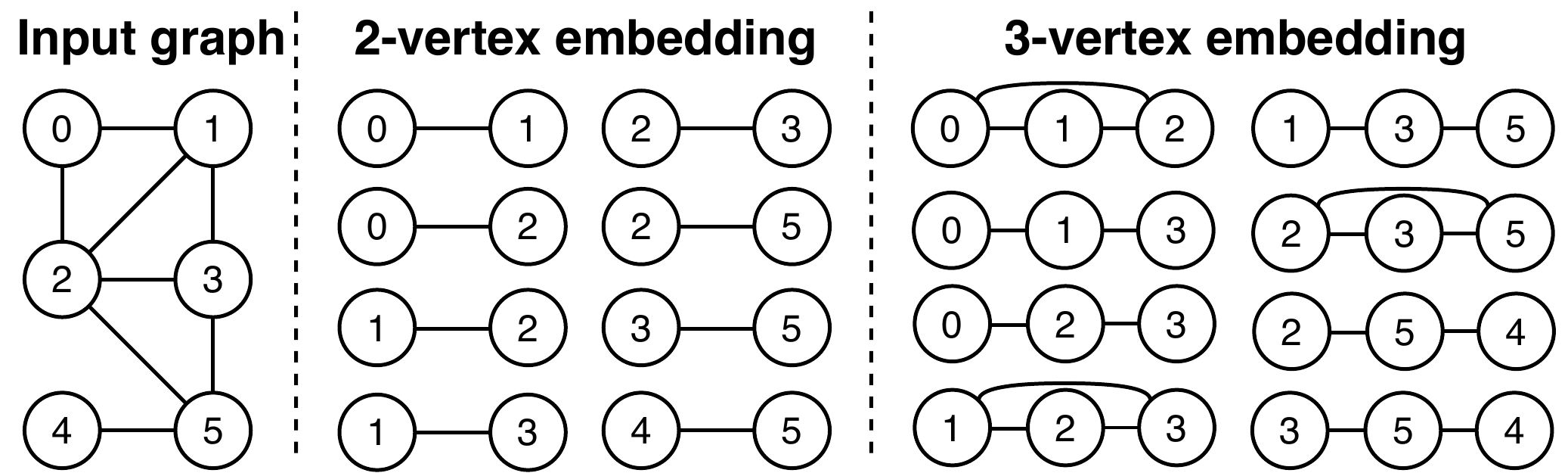}
	\vspace{-0cm}
	\caption{\small An example of vertex extension.}
	\vspace{-0.3cm}
	\label{fig:extension}
	\end{minipage}
\hfill
	\begin{minipage}[t]{0.32\linewidth}
	\centering
	\includegraphics[width=\textwidth]{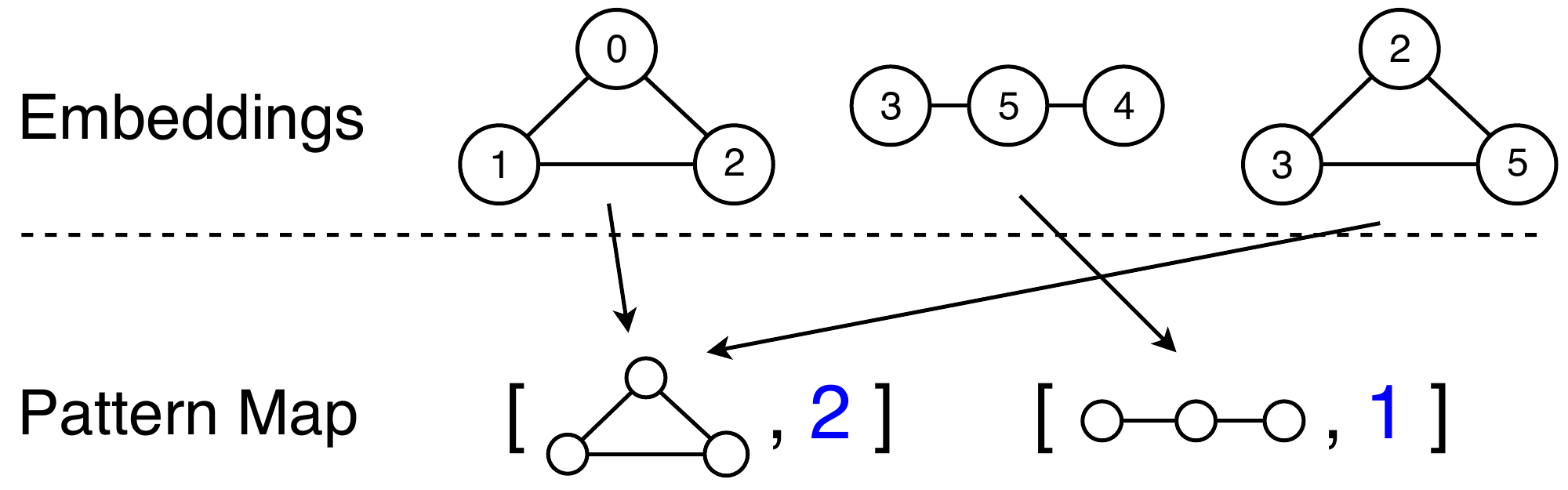}
	\vspace{-0.3cm}
	\caption{\small Reduction operation that calculates pattern
	frequency using a pattern map.}
	\vspace{-0.3cm}
	\label{fig:reduce}
	\end{minipage}
\hfill
	\begin{minipage}[t]{0.3\linewidth}
	\centering
	\includegraphics[width=\textwidth]{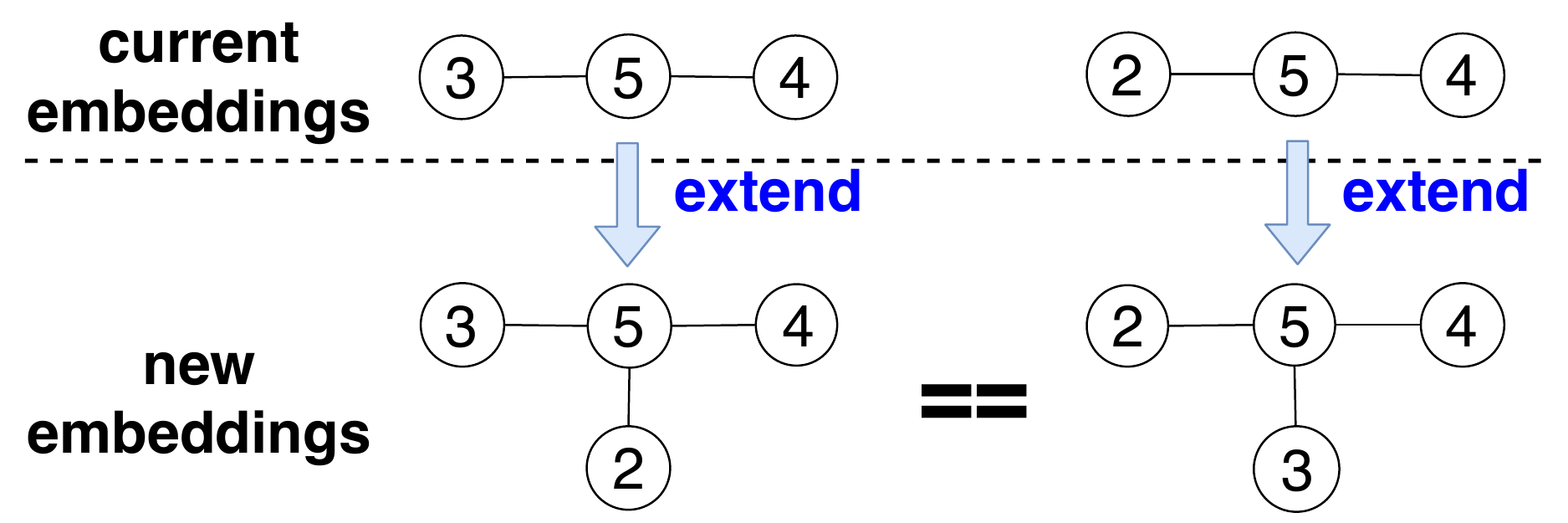}
	\vspace{-0cm}
	\caption{\small An example of automorphism.}
	\vspace{-0.3cm}
	\label{fig:automorphism}
	\end{minipage}
\end{figure*}

\subsection{Execution Model}\label{subsect:model}
Algorithm~\ref{alg:engine} describes the execution engine in Pangolin
which illustrates our {\it extend-reduce-filter} execution model. 
To begin with, a worklist of embeddings is initialized with all the
single-edge embeddings (line 4). The engine then works in an 
iterative fashion (line 6). In each iteration, i.e., \textit{level}, 
there are three phases: \textproc{Extend} (line 8), \textproc{Reduce} 
(line 10) and \textproc{Filter} (line 12). 
Pangolin exposes necessary details in each phase to enable a more 
flexible programming interface (\cref{subsect:apis}) than existing systems; 
for example, Pangolin exposes the \textproc{Extend} phase
which is implicit in Arabesque.  

The \textproc{Extend} phase takes each embedding in the input worklist 
and extends it with a vertex (vertex-induced) or an edge (edge-induced).
Newly generated embeddings then form the output worklist for the next level.
The embedding size is increased with $level$ until the user defined maximum size is reached (line 14). 
\cref{fig:extension} shows an example of the first iteration of vertex-based extension. 
The input worklist consists of all the 2-vertex (i.e., single-edge) embeddings.
For each embedding in the worklist, one vertex is added to yield a 
3-vertex embedding. For example, the first 2-vertex embedding $\{0,1\}$ is extended 
to two new 3-vertex embeddings $\{0,1,2\}$ and $\{0,1,3\}$. 

\makeatletter
\newcommand{\algorithmicbreak}{\textbf{break}}
\newcommand{\Break}{\State \algorithmicbreak}
\makeatother

\setlength{\textfloatsep}{2pt}
\begin{algorithm}[t]
\small
	\caption{Execution Model for Mining}
	\label{alg:engine}
	\begin{algorithmic}[1]
		\Procedure{MineEngine}{$G$($V$,$E$), MAX\_SIZE}
		\State EmbeddingList $in\_wl$, $out\_wl$ \Comment{double buffering}
		\State PatternMap $p\_map$
		\State \Call{Init}{$in\_wl$} \Comment{insert single-edge embeddings}
		\State $level \leftarrow 1$
		\While{true}
		\State $out\_wl \leftarrow \emptyset$ \Comment{clear the new worklist}
		\State \Call{\textcolor{blue}{Extend}}{$in\_wl$, $out\_wl$}
		\State $p\_map \leftarrow \emptyset$ \Comment{clear the pattern map}
		\State \Call{\textcolor{blue}{Reduce}}{$out\_wl$, $p\_map$}
		\State $in\_wl \leftarrow \emptyset$ \Comment{clear the old worklist}
		\State \Call{\textcolor{blue}{Filter}}{$out\_wl$, $p\_map$, $in\_wl$}
		\State $level \leftarrow level + 1$
		\If{$level =$ MAX\_SIZE - 1}
		\Break \Comment{termination condition}
		\EndIf
		\EndWhile
		\State \Return $in\_wl$, $p\_map$
		\EndProcedure
	\end{algorithmic}
\end{algorithm}

After vertex/edge extension, a \textproc{Reduce} phase is used to 
extract some pattern-based statistical information, 
i.e., pattern frequency or {\it support}, from the embedding worklist. 
The \textproc{Reduce} phase first classifies all the embeddings 
in the worklist into different categories according to their patterns, 
and then computes the support for each pattern category, 
forming pattern-support pairs. 
All the pairs together constitute a pattern map ($p\_map$ in line 10). 
\cref{fig:reduce} shows an example of the reduction operation. 
The three embeddings (top) can be classified into two categories, 
i.e., triangle and wedge (bottom). 
Within each category, this example counts the number of embeddings as the support. 
As a result, we get the pattern-map as \{[triangle, 2], [wedge, 1]\}. 
After reduction, a \textproc{Filter} phase may be needed to remove those
embeddings which the user are no longer interested in; 
e.g., FSM removes infrequent embeddings in this phase. 

Note that \textproc{Reduce} and \textproc{Filter} phases are not 
necessary for all applications, and they can be disabled by the user. 
If they are used, they are also executed after initializing 
single-edge embeddings (line 4) \hlc{and before entering the main loop (line 6). 
Thus, infrequent single-edge embeddings are filtered out 
to collect only the frequent ones 
before the main loop starts.
Note that this is omitted from Algorithm~\ref{alg:engine} due to lack of space.}
If \textproc{Reduce} is enabled but \textproc{Filter} is disabled, 
then reduction is only required and executed for the last iteration,
as the pattern map produced by reduction is not used in prior iterations (dead code). 

\setlength{\textfloatsep}{1pt}
\begin{algorithm}[t]
\small
	\caption{Compute Phases in \textbf{Vertex-induced} Mining}
	\label{alg:operators}
	\begin{algorithmic}[1]
		\Procedure{Extend}{$in\_wl$, $out\_wl$}
		\For{each embedding $emb \in in\_wl$ \textbf{in parallel}}
			\For{each vertex $v$ in $emb$}
				\If{\Call{\textcolor{blue}{toExtend}}{$emb$, $v$} = $true$}
					\For{each vertex $u$ in $adj(v)$}
						\If{\Call{\textcolor{blue}{toAdd}}{$emb$, $u$} = $true$}
							\State insert $emb \cup u$ to $out\_wl$
						\EndIf
					\EndFor
				\EndIf
			\EndFor
		\EndFor
		\EndProcedure
		\item[]
		\Procedure{Reduce}{$queue$, $p\_map$}
		\For{each embedding $emb \in queue$ \textbf{in parallel}}
			\State Pattern $pt$ $\leftarrow$ \Call{\textcolor{blue}{getPattern}}{$emb$}
			\State Support $sp$ $\leftarrow$ \Call{\textcolor{blue}{getSupport}}{$emb$}
				\State $p\_map$[$pt$] $\leftarrow$ \Call{\textcolor{blue}{Aggregate}}{$p\_map$[$pt$], $sp$}
		\EndFor
		\EndProcedure
		\item[]
		\Procedure{Filter}{$in\_wl$, $p\_map$, $out\_wl$}
		\For{each embedding $emb \in in\_wl$ \textbf{in parallel}}
			\If{\Call{\textcolor{blue}{toPrune}}{$emb$, $p\_map$} = $false$}
				\State insert $emb$ to $out\_wl$
			\EndIf
		\EndFor
		\EndProcedure
	\end{algorithmic}
\end{algorithm}

\setlength{\textfloatsep}{\parsep}

\subsection{Programming Interface}\label{subsect:apis}
Pangolin exposes flexible and simple interfaces to the user 
to express application-specific optimizations. 
\cref{lst:api} lists user-defined functions (APIs) and 
\cref{alg:operators} describes how these functions (marked in blue) 
are invoked by the Pangolin execution engine. 
A specific application can be created by defining these APIs. 
Note that all the functions are not mandatory; each of them has
a default return value.

In the \textproc{Extend} phase, we provide two functions,
\texttt{toAdd} and \texttt{toExtend},
for the user to prune embedding candidates aggressively.
When they return false, the execution engine avoids generating 
an embedding and thus the search space is reduced. 
More specifically, \texttt{toExtend} checks whether a 
vertex in the current embedding needs to be extended. 
Extended embeddings can have duplicates due to {\it automorphism}.
\cref{fig:automorphism} illustrates {\it automorphism}: two different embeddings 
$(3,5,4)$ and $(2,5,4)$ can be extended into the same embedding $(2,5,3,4)$.
Therefore, only one of them (the canonical embedding) should be kept, 
and the other (the redundant one) should be removed. 
\hlc{This is done by a \textit{canonical test} in \texttt{toAdd}, 
which checks whether the newly generated 
embedding is a \emph{qualified} candidate.
An embedding is not qualified when it is a duplicate or it does not 
have certain user-defined characteristics. 
Only qualified embeddings are added into the next worklist.}
Application-specific knowledge can be used to specialize the two functions.
If left undefined, \texttt{toExtend} returns true 
and \texttt{toAdd} does a default canonical test. 
Note that the user specifies whether the embedding exploration 
is vertex-induced or edge-induced.
The only difference for edge-induced extension is in lines 5 to 7: 
instead of vertices adjacent to $v$, edges incident on $v$ are used.


\begin{lstlisting}[float=tp,floatplacement=b,label={lst:api},language=C++,abovecaptionskip=0pt,caption={User-defined functions in Pangolin.}]
bool toExtend(Embedding emb, Vertex v);
bool toAdd(Embedding emb, Vertex u)
bool toAdd(Embedding emb, Edge e)
Pattern getPattern(Embedding emb)
Pattern getCanonicalPattern(Pattern pt)
Support getSupport(Embedding emb)
Support Aggregate(Support s1, Support s2)
bool toPrune(Embedding emb);
\end{lstlisting}

In the \textproc{Reduce} phase, \texttt{getPattern} function specifies how to obtain the pattern of an embedding. 
Finding the canonical pattern of an embedding involves an expensive {\it isomorphism} test. 
This can be specialized using application-specific knowledge 
to avoid such tests.
If left undefined, a canonical pattern is returned by 
\texttt{getPattern}. 
In this case,
to reduce the overheads of invoking the {\it isomorphism} test,
embeddings in the worklist are first reduced using their {\it quick patterns}~\cite{Arabesque},
and then quick patterns are aggregated using their canonical patterns. 
In addition, \texttt{getSupport} and {\tt Aggregate} 
functions specify the support of an embedding and the reduction 
operator for the support, respectively. 


Lastly, in the \textproc{Filter} stage, \texttt{toPrune} is used to
specify those embeddings the user is no longer interested in. 
This depends on the support for the embedding's canonical pattern 
(that is in the computed pattern map). 

\label{para:complexity}
\hlc{\textbf{Complexity Analysis.}
Consider an input graph $G$ with $n$ vertices and maximum embedding size $k$. 
In the \textproc{Extend} phase of the last level (which dominates the execution time and complexity), 
there are up to $O(n^{k-1})$ embeddings in the input worklist.
Each embedding has up to $k-1$ vertices to extend.
Each vertex has up to $d_{max}$ 
neighbors (candidates). 
In general, 
each candidate needs to check connectivity with $k-1$ vertices, 
with a complexity of $O(log(d_{max}))$ (binary search).
An isomorphism test needs to be performed 
for each newly generated embedding (size of $k$) to find its pattern.
The state-of-the-art algorithm to test isomorphism 
has a complexity of $O(e^{\sqrt{klogk}})$~\cite{Complexity}.
Therefore, the overall worst-case complexity is $O(n^{k-1}k^2d_{max}log(d_{max})e^{\sqrt{klogk}})$.}

Pangolin also provides APIs to process the embeddings or pattern 
maps at the end of each phase (e.g., this is used in clique-listing, 
which a variant of clique-finding that requires listing all the cliques). 
We omit this from Algorithm~\ref{alg:operators} 
and \cref{lst:api} for the sake of brevity.
To implement the application-specific functions,
users are required to write C++ code for CPU and 
CUDA \texttt{\_\_device\_\_} functions for GPU 
(compiler support can provide a unified interface 
for both CPU and GPU in the future). 
\cref{lst:routine} lists the helper routines provided by Pangolin. 
These routines are commonly used in GPM applications;
e.g., to check connectivity, to test canonicality, 
as well as an implementation of domain support. 
They are available on both CPU and GPU, 
with efficient implementation on each architecture. 

\textbf{Comparison With Other GPM APIs:}
Existing GPM frameworks do not expose \texttt{toExtend} 
and \texttt{getPattern} to the application developer 
(instead, they assume these functions always return 
true and a canonical pattern, respectively).
\hlc{Note that existing embedding-centric frameworks like Arabesque
can be extended to expose the same API functions in Pangolin so as to 
enable application-specific optimizations (Section~\ref{sect:app-opt}),
but this is difficult for relational model based systems 
like RStream, as the table join operations
are inflexible to allow this fine-grained control.}


\begin{lstlisting}[float=tp,floatplacement=tbp,label={lst:routine},language=C++,abovecaptionskip=0pt,caption={Helper routines provided to the user by Pangolin.}]
// connectivity checking routines
bool isConnected(Vertex u, Vertex v)

// canonical test routines
bool isAutoCanonical(Embedding emb, Vertex v)
bool isAutoCanonical(Embedding emb, Edge e)
Pattern getIsoCanonicalBliss(Embedding emb)
Pattern getIsoCanonicalEigen(Embedding emb)

// to get domain (MNI) support
Support getDomainSupport(Embedding emb)
Support mergeDomainSupport(Support s1, Support s2)
Support getPatternSupport(Embedding emb)
Support getPatternSupport(Edge e)
\end{lstlisting}

\begin{lstlisting}[float=tp,floatplacement=tbp,label={lst:kcl},language=C++,abovecaptionskip=0pt,caption={\small Clique finding (vertex induced) in Pangolin.}]
bool toExtend(Embedding emb, Vertex v) {
	return (emb.getLastVertex() == v); 
}
bool toAdd(Embedding emb, Vertex u) {
	for v in emb.getVertices() except last:
		if (!isConnected(v, u)) return false;
	return true; 
}
\end{lstlisting}

\subsection{Applications in Pangolin}\label{subsect:apps}
TC, CF, and MC use vertex-induced embeddings, while FSM uses edge-induced embeddings. 
\cref{lst:kcl,lst:motif,lst:fsm} show CF, MC, and FSM implemented in Pangolin
(we omit TC due to lack of space). 
For TC, extension happens only once, i.e., for each edge $(v_0, v_1$),
$v_1$ is extended to get a neighbor $v_2$. 
We only need to check whether $v_2$ is connected to $v_0$.
If it is, this 3-vertex embedding $(v_0, v_1, v_2)$ forms a triangle. 
For CF in \cref{lst:kcl}, the search space is reduced by extending only 
the last vertex in the embedding instead of extending every vertex. 
If the newly added vertex is connected to all the vertices in the embedding,
the new embedding forms a clique. Since cliques can only grow
from smaller cliques (e.g., 4-cliques can only be generated by extending
3-cliques), all the non-clique embeddings are implicitly pruned. 
Both TC and CF do not use \textproc{Reduce} and \textproc{Filter} phases.


\begin{lstlisting}[float=tp,floatplacement=tbp,label={lst:motif},language=C++,abovecaptionskip=0pt,caption={\small Motif counting (vertex induced) in Pangolin.}]
bool toAdd(Embedding emb, Vertex v) {
	return isAutoCanonical(emb, v); 
}
Support getSupport(Embedding emb) { return 1; }
Pattern getPattern(Embedding emb) {
	return getIsoCanonicalBliss(emb); 
}
Support Aggregate(Support s1, Support s2) {
	return s1 + s2; 
}
\end{lstlisting}

\cref{lst:motif} shows MC. An extended embedding is added only if it is 
canonical according to automorphism test. In the \textproc{Reduce} phase,
the quick pattern of each embedding is first obtained
and then the canonical pattern is obtained using an isomorphism test.
In Section~\ref{subsect:hybrid}, we show a way to customize this pattern
classification method for MC to improve performance. \textproc{Filter} 
phase is not used by MC. 

\begin{lstlisting}[float=tp,floatplacement=tbp,label={lst:fsm},language=C++,abovecaptionskip=0pt,caption={\small Frequent subgraph mining (edge induced) in Pangolin.}]
bool toAdd(Embedding emb, Edge e) {
	return isAutoCanonical(emb,e) 
}
Support getSupport(Embedding emb) {
	return getDomainSupport(emb); 
}
Pattern getCanonicalPattern(Embedding emb) {
	return getIsoCanonicalBliss(emb); 
}
Support Aggregate(Support s1, Support s2) {
	return mergeDomainSupport(s1, s2); 
}
bool toPrune(Embedding emb, PatternMap map) {
	return (getPatternSupport(emb, map) < MIN_SUPPORT)
}
\end{lstlisting}

FSM is the most complicated GPM application. 
As shown in \cref{lst:fsm}, it uses the custom domain support 
routines provided by Pangolin. An extended embedding is added only if 
the new embedding is (automorphism) canonical. FSM uses the \textproc{Filter} 
phase to remove embeddings whose patterns are not frequent from the worklist. 
Despite the complexity of FSM, the Pangolin implementation is still much 
simpler than hand-optimized FSM implementations~\cite{DistGraph,Scalemine,GraMi},
thanks to the Pangolin API and helper routines.

\section{Supporting Application-Specific Optimizations in Pangolin}\label{sect:app-opt}

In this section, we describe how Pangolin's API and execution 
model supports application-specific optimizations that:
(1) enable enumeration search space pruning 
and (2) enable the eliding of isomorphism tests.

\subsection{Pruning Enumeration Search Space}

\textbf{Directed Acyclic Graph (DAG):}
In typical GPM applications, 
the input graph is undirected.
In some vertex-induced GPM applications, a common optimization
technique is \textit{orientation} which converts the undirected
input graph into a directed acyclic graph (DAG)~\cite{Arboricity,Alon}.
Instead of enumerating candidate subgraphs in an undirected graph,
the direction significantly cuts down the combinatorial search
space. Orientation has been adopted in triangle
counting~\cite{Schank}, clique finding~\cite{KClique},
and motif counting~\cite{ESCAPE}. \cref{fig:dag} illustrates an example of the
DAG construction process. In this example, vertices are ordered
by vertex ID. Edges are directed from vertices with smaller IDs
to vertices with larger IDs. Generally, vertices can be ordered
in any total ordering, which guarantees the input graph is
converted into a DAG. In our current implementation, we establish
the order~\cite{DistTC} among the vertices based on their degrees: edges will
point towards the vertex with higher degree. When there is a tie,
the edge points to the vertex with the larger vertex ID.
Other orderings can be included in the future.
In Pangolin, the user can enable orientation by simply setting a macro.

\textbf{Eager Pruning:}
In some applications like MC and FSM, 
all vertices in an embedding may need to be extended 
before determining whether the new embedding candidate 
is a ({\it automorphism}) canonical embedding or a duplicate.
However, in some applications like TC and CF~\cite{KClique}, 
duplicate embeddings can be detected 
eagerly before extending current embeddings. 
In both TC and CF, 
all embeddings obtained by extending vertices 
except (the last) one will lead to duplicate embeddings.
Thus, as shown in Listing~\ref{lst:kcl}, 
only the last vertex of the current embedding needs to be extended.
This aggressive pruning can significantly reduce the search space.
The \texttt{toExtend} function in Pangolin 
enables the user to specify such {\it eager pruning}. 

\subsection{Eliding Isomorphism Test}\label{subsect:hybrid}

\begin{figure}[t]
\centering
	\begin{minipage}[t]{0.45\linewidth}
	\centering
	\includegraphics[width=\textwidth]{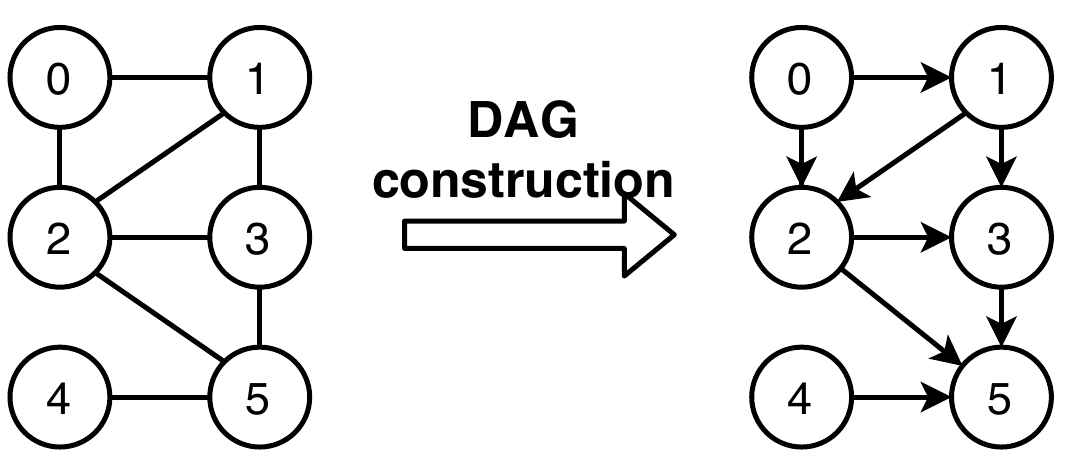}
	\vspace{-0.27cm}
	\caption{\small Convert an undirected graph into a DAG.}
	\label{fig:dag}
	\end{minipage}
\hfill
	\begin{minipage}[t]{0.49\linewidth}
	\centering
	\includegraphics[width=\textwidth]{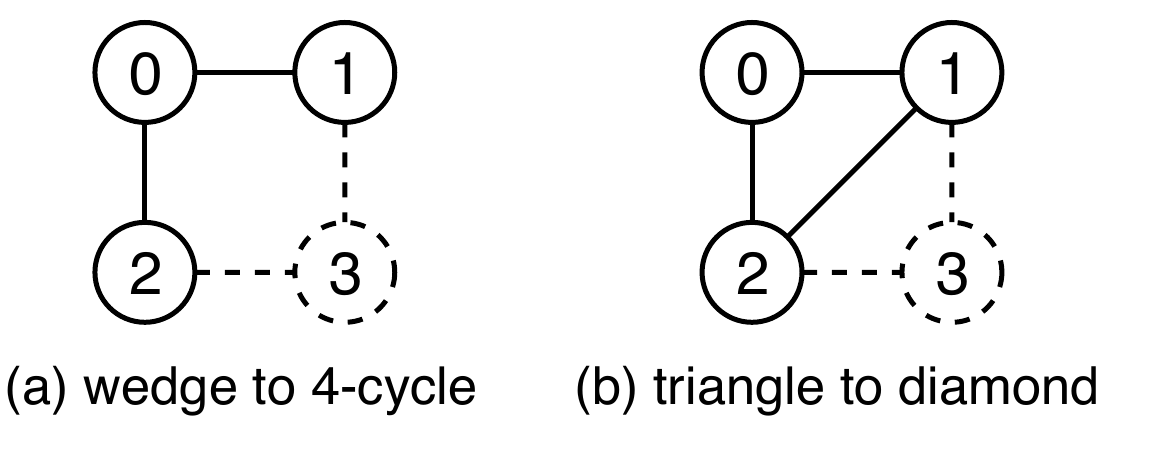}
	\vspace{-0.27cm}
	\caption{\small Examples of eliding isomorphism test for 4-MC.}
	\label{fig:avoid-iso}
	\end{minipage}
	\vspace{-0.1cm}
\end{figure}

\textbf{Exploiting Memoization:}
Pangolin avoids redundant computation in each stage with memoization.
Memoization is a tradeoff between computation and memory usage. 
Since GPM applications are usually memory hungry,
we only do memoization when it requires small amount of 
memory and/or it dramatically reduce complexity.
For example, in the \textproc{Filter} phase of FSM, Pangolin avoids isomorphism
test to get the pattern of each embedding, since it has been done in the
\textproc{Reduce} phase. This recomputation is avoided by maintaining a pattern 
ID (hash value) in each embedding after isomorphism test, and setting up
a map between the pattern ID and pattern support. Compared to isomorphism test,
which is extremely compute and memory intensive, storing the pattern 
ID and a small pattern support map is relatively lightweight.
In MC, which is another application to find multiple patterns,
the user can easily enable memoization for the pattern id in each level.
In this case, when it goes to the next level, the pattern of each 
embedding can be identified with its pattern id in the previous level 
with much less computation than a generic isomorphism test. 
As shown in \cref{fig:avoid-iso}, to identify a 4-cycle
from a wedge or a diamond from a triangle, we only need to check if
vertex 3 is connected to both vertex 1 and 2. 

\begin{lstlisting}[float=t,floatplacement=b,label={lst:3-motif},language=C++,abovecaptionskip=0pt,belowcaptionskip=2pt,caption=Customized pattern classification for 3-MC.]
Pattern getPattern(Embedding emb) { 
	if (emb.size() == 3) {
		if (emb.getNumEdges() == 3) return P1;
		else return P0;
	} else return getIsoCanonicalBliss(emb);
}
\end{lstlisting}

\textbf{Customized Pattern Classification:}
In the \textproc{Reduce} phase, embeddings are classified into different 
categories based on their patterns, as shown in \cref{fig:reduce}.
To get the pattern of an embedding, a generic way is to convert 
the embedding into a canonical graph that is isomorphic to it 
(done in two steps, as explained in Section~\ref{subsect:apis}). 
Like Arabesque and Rstream, Pangolin uses the Bliss~\cite{Bliss} 
library for getting the canonical graph or pattern for an embedding.
This graph isomorphism approach is applicable to embeddings of any size, 
but it is very expensive as it requires frequent dynamic memory 
allocation and consumes a huge amount of memory. For small embeddings, 
such as 3-vertex and 4-vertex embeddings in vertex-induced applications 
and 2-edge and 3-edge embeddings in edge-induced applications, 
the canonical graph or pattern can be computed very efficiently.
For example, we know that there are only 2 patterns in 3-MC
(i.e., wedge and triangle in \cref{fig:4-motifs}).
The only computation needed to differentiate the two patterns is to 
count the number of edges (i.e., a wedge has 2 edges and a triangle has 3), 
as shown in \cref{lst:3-motif}. This specialized method significantly 
reduces the computational complexity of pattern classification. 
The \texttt{getPattern} function in Pangolin enables the user 
to specify such \textit{customized pattern classification}.

\section{Implementation on CPU and GPU}\label{sect:impl}

The user implements application-specific optimizations 
using the Pangolin API and helper functions, 
and Pangolin transparently parallelizes the application. 
Pangolin provides an efficient and scalable parallel 
implementation on both shared-memory multicore CPU and GPU. 
Its CPU implementation is built using 
the Galois~\cite{Galois} libray and 
its GPU implementation is built using 
the LonestarGPU~\cite{LonestarGPU} infrastructure. 
Pangolin includes several architectural optimizations. 
In this section, we briefly describe some of them:
(1) exploiting locality and fully utilizing memory 
bandwidth~\cite{Analysis,Locality,Memory}; 
(2) reducing the memory consumption;
(3) mitigating the overhead of dynamic memory allocation;
(4) minimizing synchronization and other overheads.

\subsection{Data Structures for Embeddings}

Since the number of possible $k$-embeddings in a graph increases
exponentially with $k$, storage for embeddings grows rapidly and 
easily becomes the performance bottleneck. Most existing systems 
use array-of-structures (AoS) to organize the embeddings, 
which leads to poor locality, especially for GPU computing.
In Pangolin, we use structure of arrays (SoA) to store embeddings in memory. 
The SoA layout is particularly beneficial for parallel processing 
on GPU as memory accesses to the embeddings are fully coalesced.

\begin{figure}[t]
\begin{center}
	\includegraphics[width=0.43\textwidth]{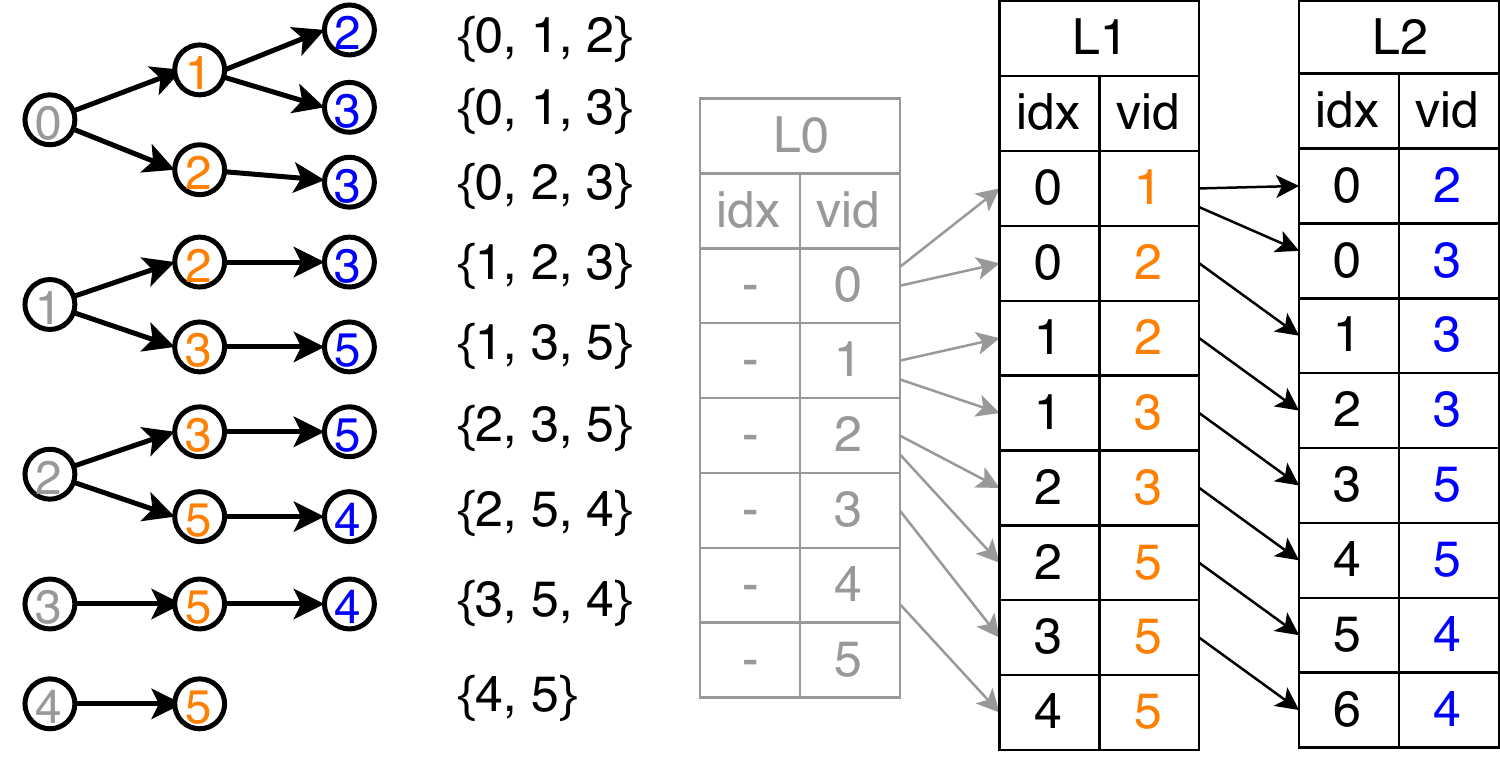}
	\vspace{-0.1cm}
	\caption{{\small An example of the embedding list data structure.}}
	\vspace{-0.47cm}
	\label{fig:embedding}
\end{center}
\end{figure}

\cref{fig:embedding} illustrates the embedding list data structure. 
On the left is the prefix-tree that illustrates the 
embedding extension process in \cref{fig:extension}. 
The numbers in the vertices are vertex IDs (VIDs). 
Orange VIDs are in the first level $L_{1}$, and blue VIDs 
belong to the second level $L_{2}$. The grey level $L_{0}$
is a dummy level which does not actually exist but is used to
explain the key ideas. On the right, we show the corresponding
storage of this prefix tree. For simplicity, we only show
the vertex-induced case. Given the maximum size $k$, 
the embedding list contains $k-1$ levels. In each level, 
there are two arrays, index array (\texttt{idx}) and vertex ID
array (\texttt{vid}). In the same position of the two arrays,
an element of index and vertex ID consists of a pair (\texttt{idx},
\texttt{vid}). In level $L_{i}$, \texttt{idx} is the index pointing
to the vertex of the same embedding in the previous level $L_{i-1}$,
and \texttt{vid} is the $i$-th vertex ID of the embedding.

Each embedding can be reconstructed by backtracking from the
last level lists. For example, to get the first embedding
in level $L_{2}$, which is a vertex set of $\{0, 1, 2\}$,
we use an empty vertex set at the beginning. We start from
the first entry ($0$, $2$) in $L_{2}$, which indicates the last
vertex ID is `$2$' and the previous vertex is at the position
of `$0$'. We put `$2$' into the vertex set $\{2\}$. Then we go back
to the previous level $L_{1}$, and get the $0$-th entry ($0$, $1$).
Now we put `$1$' into the vertex set $\{1, 2\}$. Since $L_{1}$ is 
the lowest level and its index is the same as the vertex ID in
level $L_{0}$, we put `$0$' into the vertex set $\{0, 1, 2\}$.

For the edge-induced case, the strategy is similar but requires 
one more column \texttt{his} in each level to indicate the history
information. Each entry is a triplet (\texttt{vid}, \texttt{his},
\texttt{idx}) that represents an edge instead of a vertex,
where \texttt{his} indicates at which level the source
vertex of this edge is, while \texttt{vid} is the ID of the destination
vertex. In this way we can backtrack the source vertex with \texttt{his}
and reconstruct the edge connectivity inside the embedding. Note that
we use three distinct arrays for \texttt{vid}, \texttt{his} and
\texttt{idx}, which is also an SoA layout. This data layout
can improve temporal locality with more data reuse. For example, the 
first \texttt{vid} in $L_{1}$ ($v_{1}$) is connected to two vertices in
$L_{2}$ ($v_{2}$ \& $v_{3}$). Therefore $v_{1}$ will be reused. Considering
high-degree vertices in power-law graphs, there are lots of reuse opportunities. 

\subsection{Avoiding Materializaton of Data Structures}
\label{sect:avoid-mat}

\textbf{Loop Fusion:}
Existing GPM systems first collect all the embedding candidates 
into a list and then call the user-defined function (like 
\texttt{toAdd}) to select embeddings from the list. 
This leads to materializaton of the candidate embeddings list. 
In contrast, Pangolin preemptively discards embedding candidates 
using the \texttt{toAdd} function before adding it to the embedding 
list (as shown in Algorithm~\ref{alg:operators}), 
thereby avoiding the materialization of the candidate embeddings 
(this is similar to {\it loop fusion} in array languages).
This significantly reduces memory allocations, 
yielding lower memory usage and execution time.

\begin{figure}[t]
\centering
	\begin{minipage}[t]{0.43\linewidth}
	\centering
	\includegraphics[width=\textwidth]{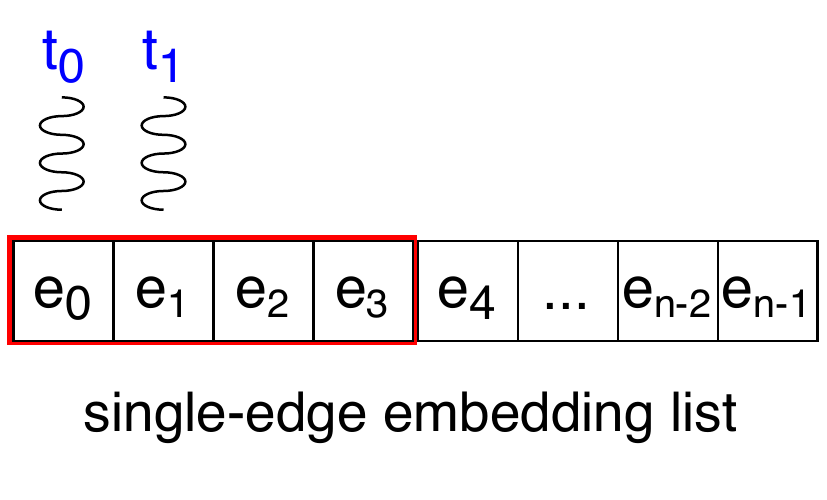}
	\vspace{-0.36cm}
	\caption{\small Edge blocking.}
	\label{fig:blocking}
	\end{minipage}
\hfill
	\begin{minipage}[t]{0.53\linewidth}
	\centering
	\includegraphics[width=\textwidth]{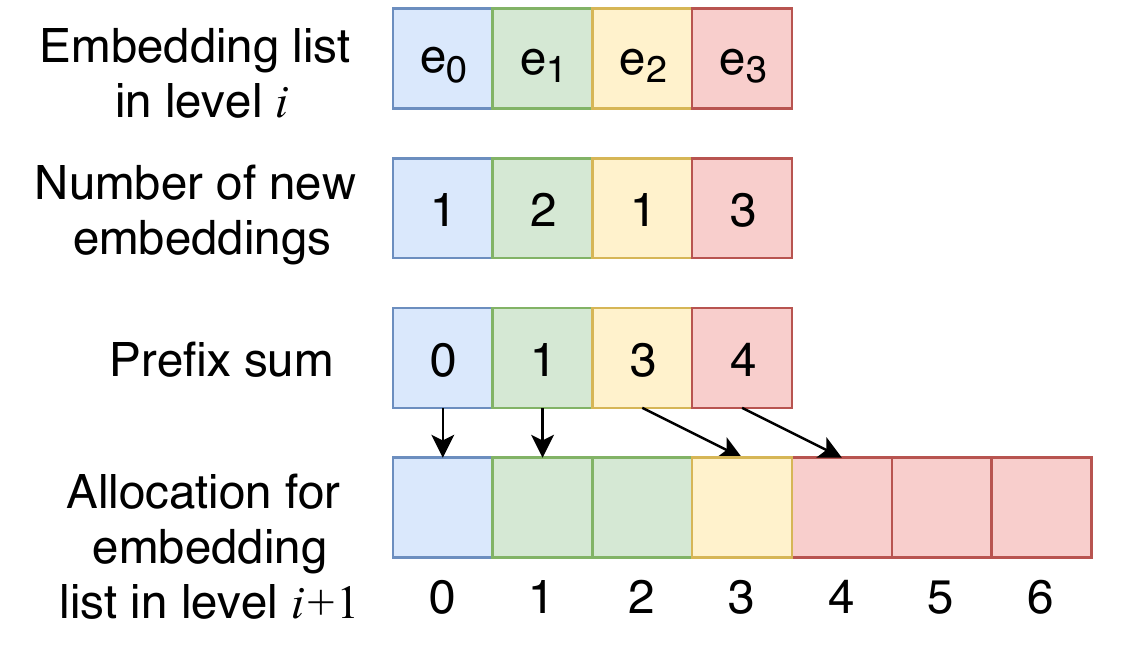}
	\vspace{-0.36cm}
	\caption{\small Inspection-execution.}
	\label{fig:insp}
	\end{minipage}
	\vspace{-0.1cm}
\end{figure}

\textbf{Blocking Schedule:}
Since the memory consumption increases exponentially with the 
embedding size, existing systems utilize either distributed
memory or disk to hold the data. However, Pangolin is a shared 
memory framework and could run out of memory for large graphs.
In order to support processing large datasets, we introduce 
an \emph{edge-blocking} technique in Pangolin.
\hlc{Since an application starts expansion with single-edge 
embeddings, Pangolin blocks the initial embedding list
into smaller chunks, and 
processes all levels (main loop in Algorithm~\ref{alg:engine}) 
for each chunk one after another.
As shown in \cref{fig:blocking}, there are $n$ edges in 
the initial embedding list ($e_0$ $\sim$ $e_{n-1}$). 
Each chunk contains 4 edges which are assigned 
to the 2 threads ($t_0$ $\sim$ $t_1$) to process.
After all levels of the current chunk are processed, 
the threads move to the next chunk and 
continue processing until all chunks are processed.
The chunk size $C_s$ is a parameter to tune;
$C_s$ is typically much larger than the number of threads.
Blocking will not affect parallelism 
because there are a large number of edges in each chunk that
can be processed concurrently. 
Note that 
the FILTER phase requires strict synchronization
in each level, 
so edge-blocking cannot be applied for applications 
that use it. For example, we need to gather embeddings for 
each pattern in FSM in order to compute the domain support.
Due to this, all embeddings needs to be processed before
moving to the next level, so we disable blocking for FSM. 
Currently, edge-blocking is used specifically for bounding
memory usage, but it is also potentially beneficial
for data locality with an appropriate block size.
We leave this for future work.}

\subsection{Dynamic Memory Allocation}
\label{sect:mem-alloc}

\textbf{Inspection-Execution:}
Compared to graph analytics applications, GPM applications
need significantly more dynamic memory allocations and  
memory allocation could become a performance bottleneck.
A major source of memory allocation is the embedding list.
As the size of embedding list increases, 
we need to allocate memory for the embeddings in each round. 
When generating the embedding list, 
there are write conflicts as different threads write to the same 
shared embedding list. In order to avoid frequent \texttt{resize} 
and \texttt{insert} operation, we use {\it inspection-execution} 
technique to generate the embedding list. 

The generation include 3 steps. In the first step,
we only calculate the number of newly generated embeddings for each 
embedding in the current embedding list. We then use parallel prefix 
sum to calculate the \texttt{start} index for each current embedding, 
and allocate the exact amount of memory for all the new embeddings. 
Finally, we actually write the new embeddings to update the embedding list, 
according to the \texttt{start} indices. In this way, each thread can 
write to the shared embedding list simultaneously without conflicts. 
\hlc{\cref{fig:insp} illustrates the inspection process.
At level $i$, there are 4 embeddings $e_0$, $e_1$, $e_2$, $e_3$ in the 
embedding list, which will generate 1, 2, 1, 3 new embeddings respectively. 
We get the \texttt{start} indices (0, 1, 3, 4) using prefix sum, 
and then allocate memory for the level $i+1$ embedding list. 
Next, each embedding writes generated embeddings from its \texttt{start} index in the level $i+1$ list (concurrently).
}

Although inspection-execution requires iterating over the embeddings twice,
making this tradeoff for GPU is reasonable for two reasons. 
First, it is fine for the GPU to do the recomputation 
as it has a lot of computation power. 
Second, improving the memory access pattern to better 
utilize memory bandwidth is more important for GPU. 
This is also a more scalable design choice for the CPU as 
the number of cores on the CPU are increasing. 

\textbf{Scalable Allocators:}
Pattern reduction in FSM is another case where dynamic memory allocation 
is frequently invoked. To compute the domain support of each pattern, 
we need to gather all the embeddings associated with the same pattern (see \cref{fig:example}). 
This gathering requires resizing the vertex set of each domain. 
The C++ standard {\tt std} library employs a concurrent allocator 
implemented by using a global lock for each allocation, 
which could seriously limit performance and scalability. 
We leverage the Galois memory allocator to alleviate this overhead. 
Galois provides an in-built efficient and concurrent memory allocator that 
implements ideas from prior scalable allocators~\cite{Hoard,Michael,Schneider}. 
The allocator uses per-thread memory pools of huge pages. 
Each thread manages its own memory pool. 
If a thread has no more space in its memory pool, 
it uses a global lock to add another huge page to its pool.
Most allocations thus avoid locks.
Pangolin uses variants of {\tt std} data structures 
provided by Galois that use the Galois memory allocator. 
\hlc{For example, this is used for maintaining the pattern map.}
On the other hand, our GPU infrastructure currently lacks support 
for efficient dynamic memory allocation inside CUDA kernels. 
To avoid frequent \texttt{resize} operations inside kernels, 
we conservatively calculate the memory space 
required and pre-allocate bit vectors for kernel use. 
This pre-allocation requires much more memory than is actually required, 
and restricts our GPU implementation to smaller inputs for FSM.

\begin{table}[t]
	\footnotesize
	\centering
	\resizebox{0.5\textwidth}{!}{
		\begin{tabular}{crrrrr}
			\Xhline{2\arrayrulewidth}
			\bf{Graph} & \bf{Source} & \multicolumn{1}{c}{\bf{\# V}} & \multicolumn{1}{c}{\bf{\# E}} & \multicolumn{1}{c}{\bf{$\overline{d}$}} & \bf{Labels}\\
			\hline
			\texttt{Mi} & {Mico}~\cite{GraMi} & 100,000 & 2,160,312 & 22 & 29\\
			\texttt{Pa} & {Patents}~\cite{Patent} & 2,745,761 & 27,930,818 & 10 & 37\\
			\texttt{Yo} & {Youtube}~\cite{Youtube} & 7,066,392 & 114,190,484 & 16 & 29\\
			\texttt{Pdb}& {ProteinDB}~\cite{DistGraph}  & 48,748,701 & 387,730,070 & 8 & 25\\
			\texttt{Lj} & {LiveJournal}~\cite{SNAP} & 4,847,571 & 85,702,474 & 18 & 0\\
			\texttt{Or} & {Orkut}~\cite{SNAP} & 3,072,441 & 234,370,166 & 76 & 0\\
			\texttt{Tw} & {Twitter}~\cite{Konect} & 21,297,772 & 530,051,090 & 25 & 0\\
			\texttt{Gsh}& {Gsh-2015}~\cite{gsh2015} & 988,490,691 & 51,381,410,236 & 52 & 0\\
			\Xhline{2\arrayrulewidth}
		\end{tabular}
	}
	\vspace{0.1cm}
	\caption{\small Input graphs (symmetric, no loops, no duplicate edges) and their properties ($\overline{d}$ is the average degree).}
	\vspace{-0.2cm}
	\label{tab:input}
\end{table}

\subsection{Other Optimizations}
\label{sect:other-opt}

\begin{figure}[t]
\begin{center}
	\includegraphics[width=0.49\textwidth]{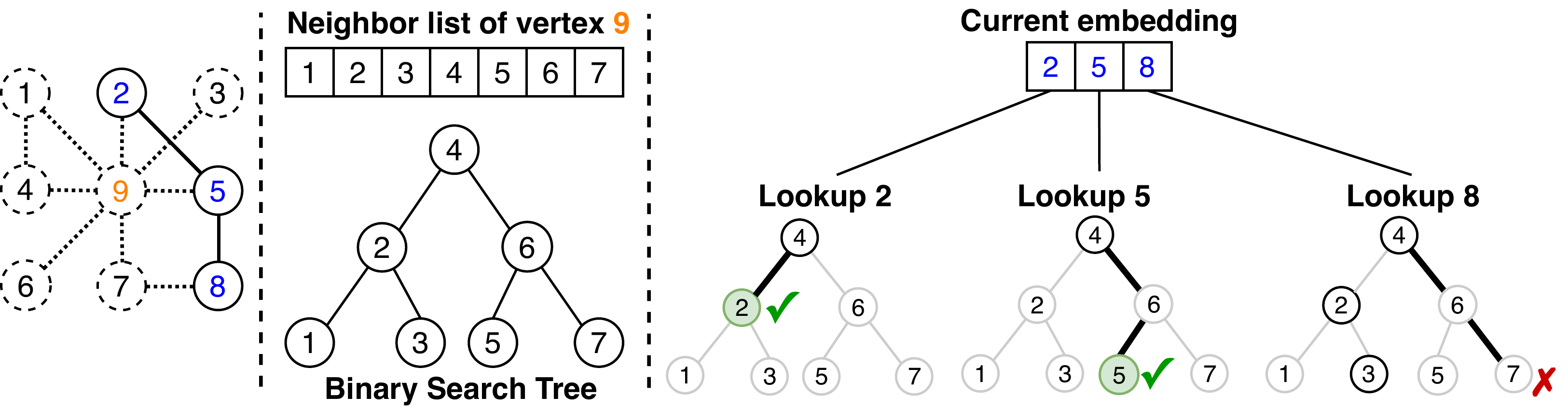}
	\caption{{\small Binary search for connectivity check.
	The current embedding is $\{2, 5, 8\}$, which
	we are trying to extend by adding vertex 9.
	To check which vertices in the embedding are directly
	connected to vertex 9, we use `2', `5', and `8' as the keys
	to search in the neighbor list of vertex 9. Using binary
	search we find `2' and `5' are connected to `9'.}}
	\vspace{-0.5cm}
	\label{fig:binary}
\end{center}
\end{figure}

\begin{table*}[t]
\centering
\small
\resizebox{\textwidth}{!}{
\begin{tabular}{c|c|rrrrr|rrrrr|rrrr}
\Xhline{3\arrayrulewidth}
\multicolumn{2}{c|}{\textbf{}} & \multicolumn{5}{c|}{\texttt{\bf Mi}} & \multicolumn{5}{c|}{\texttt{\bf Pa}} & \multicolumn{4}{c}{\texttt{\bf Yo}} \\ \hline
\textbf{App} & \textbf{Option} & \bf{AR} & \bf{RS} & \bf{KA$^\dagger$} & \bf{FR} & \bf{PA} & \bf{AR} & \bf{RS} & \bf{KA$^\dagger$} & \bf{FR} & \bf{PA} & \bf{AR} & \bf{RS} & \bf{KA$^\dagger$} & \bf{PA} \\ \hline
\textbf{TC}        &   & 30.8 & 2.6 & 0.2 & & \textbf{0.02} & 100.8 & 7.8 & 0.5 && \textbf{0.08} & 601.3 & 39.8 & 2.2& \textbf{0.3} \\ \hline
\multirow{3}{*}{\textbf{CF}} & 3 & 32.2 & 7.3 & 0.5 & 24.7 & \textbf{0.04}& 97.8 & 39.1 & 0.6 & 350.2 & \textbf{0.2} & 617.0 & 862.3 & 2.2 & \textbf{0.7} \\
                   & 4 & 41.7 & 637.8 & 3.9 & 30.6 & \textbf{1.6} & 108.1 & 62.1 & 1.1 & 410.1 & \textbf{0.4} & 1086.9 & - & 7.8 & \textbf{3.1} \\
                   & 5 & 311.9& - & 183.6 & 488.9  &\textbf{60.5} & 108.8 & 76.9 & 1.5 & 463.5 & \textbf{0.5} & 1123.6 & - & 19.0& \textbf{7.3} \\ \hline
\multirow{2}{*}{\textbf{MC}} & 3 & 36.1 & 7137.5 & 1.4 & 41.2 & \textbf{0.2} & 101.6 &3886.9 & 4.7 & 236.3 & \textbf{0.9} & 538.4 & 89387.0 & 35.5 & \textbf{5.5} \\
                   & 4 & 353.0 & - & 198.2 & 243.2 &\textbf{175.6} & 779.8 & - & \textbf{152.3} & 561.1 & 209.1 & 5132.8& - & 4989.0 & \textbf{4405.3}\\ \hline
\multirow{4}{*}{\textbf{3-FSM}}& 300 & 104.9 & 56.8 & 7.4 & 780.5 &\textbf{3.9} & 340.7 & 230.1 & 25.5 & 720.3 & \textbf{14.7} & 666.9 & 1415.1 & 132.6 & \textbf{96.9} \\
                   & 500 & 72.2 & 57.9 & 8.2 & 773.1 & \textbf{3.6} & 433.6 & 208.6 & 26.4 & 817.0 & \textbf{15.8} & 576.5 & 1083.9 & 133.3 & \textbf{97.8}  \\
                   & 1000& 48.5 & 52.9 & 7.8 & 697.2 & \textbf{3.0} & 347.3 & 194.0 & 28.7 & 819.9 & \textbf{18.1} & 693.2 & 1179.3 & 136.2 & \textbf{98.0}  \\
                   & 5000& 36.4 & 35.6 & 3.9 & 396.3 & \textbf{2.4} & 366.1 & 172.2 & 31.5 & 915.5 & \textbf{27.0} & 758.6 & 1248.1 & 155.0 & \textbf{102.2} \\
\Xhline{3\arrayrulewidth}
\end{tabular}
}
\vspace{0.07cm}
\caption{\small Execution time (sec) of applications in GPM frameworks on 28 cores
(option: minimum support for 3-FSM; $k$ for others).
AR, RS, KA, FR, and PA: Arabesque, RStream, Kaleido, Fractal, and Pangolin respectively.
`-': out of memory or disk, or timed out in 30 hours.
FR for \texttt{Yo} is omitted due to failed execution.
FR does not contain TC. 
$^\dagger$KA results are reported from their paper.}
\label{tab:compare}
\vspace{-0.5cm}
\end{table*}

GPM algorithms make extensive use of connectivity operations
for determining how vertices are connected in the input graph. For 
example, in $k$-cliques, we need to check whether a new vertex is 
connected to all the vertices in the current embedding. Another common 
connectivity operation is to determine how many vertices are connected 
to given vertices $v_{0}$ and $v_{1}$, which is usually obtained by 
computing the intersection of the neighbor lists of the two vertices.
A naive solution of connectivity checking is to search for one vertex 
$v_{0}$ in the other vertex $v_{1}$'s neighbor list sequentially. 
If found, the two vertices are directly connected. 
To reduce complexity and improve parallel efficiency, 
we employ binary search for the connectivity check. 
\hlc{\cref{fig:binary} illustrates an example of 
binary search for connectivity check.}
This is particularly efficient on GPU, 
as it improves GPU memory efficiency~\cite{TriCore}.
We provide efficient CPU and GPU implementations of 
these connectivity operations as helper routines,
such as \texttt{isConnected} (\cref{lst:routine}), which allow 
the user to easily compose pruning strategies in applications.


\hlc{In summary, when no algorithmic optimization is applied,
programming in Pangolin should be as easy as previous GPM systems like Arabesque.
In this case, performance gains over Arabesque is achieved
due to the architectural optimizations (e.g., data structures) in Pangolin.
To incorporate algorithmic optimizations, the user can leverage
Pangolin API functions (e.g., \texttt{toExtend} and \texttt{toAdd}) to express
application-specific knowledge. 
While this involves slightly more programming effort, 
the user can get an order of magnitude performance improvement by doing so.}

\section{Evaluation}
\label{sect:eval}

In this section, 
we compare Pangolin with state-of-art GPM frameworks
and hand-optimized applications.
We also analyze Pangolin performance in more detail.

\subsection{Experimental Setup}\label{subsect:setup}

\hlc{We compare Pangolin with the state-of-the-art GPM frameworks: 
Arabesque~\cite{Arabesque}, RStream~\cite{RStream}, 
G-Miner~\cite{G-Miner}, Kaleido~\cite{Kaleido}, 
Fractal~\cite{Fractal}, and AutoMine~\cite{AutoMine}. 
Arabesque, G-Miner, and Fractal support distributed execution, 
while the rest support out-of-core execution. 
All of them support execution only on CPUs.
Kaleido and AutoMine results are reported 
from their papers because they are not publicly available.
We also compare Pangolin with the state-of-the-art 
hand-optimized GPM applications~\cite{GAPBS,DistTC,KClique,PGD,Rossi,DistGraph,ParFSM,GpuFSM}.}

We test the 4 GPM applications discussed in Section~\ref{subsect:apps},
i.e., TC, CF, MC, and FSM.
$k$-MC and $k$-CF terminate when subgraphs reach a size of $k$ vertices.
For $k$-FSM, we mine the frequent subgraphs with $k-1$ edges.
\cref{tab:input} lists the input graphs used in the experiments.
We assume that input graphs are symmetric, have no self-loops,
and have no duplicated edges. We represent the input graphs in 
memory in a compressed sparse row (CSR) format. The neighbor 
list of each vertex is sorted by ascending vertex ID.

The first 3 graphs --- \texttt{Mi}, \texttt{Pa}, and \texttt{Yo} --- have been 
previously used by Arabesque, RStream, and Kaleido. We use the same
graphs to compare Pangolin with these existing frameworks.
In addition, we include larger graphs from SNAP Collection~\cite{SNAP} 
(\texttt{Lj}, \texttt{Or}), Koblenz Network Collection~\cite{Konect} (\texttt{Tw}), 
DistGraph~\cite{DistGraph}(\texttt{Pdb}), and a very large web-crawl~\cite{gsh2015} (\texttt{Gsh}).
Except \texttt{Pdb}, other larger graphs do not have 
vertex labels, therefore, we only use them to test TC, CF, and MC.
\texttt{Pdb} is used only for FSM.

Unless specified otherwise, CPU experiments were conducted on a single machine with 
Intel Xeon Gold 5120 CPU 2.2GHz, 4 sockets (14 cores each), 
190GB memory, and 3TB SSD.
\hlc{AutoMine was evaluated using 40 threads 
(with hyperthreading) on
Intel Xeon E5-2630 v4 CPU 2.2GHz, 2 sockets (10 cores each), 
64GB of memory, and 2TB of SSD.}
Kaleido was tested using 56 threads (with hyperthreading) on 
Intel Xeon Gold 5117 CPU 2.0GHz, 2 sockets (14 cores each), 128GB memory, and 480GB SSD.
To make our comparison fair, we restrict our experiments to 
use only 2 sockets of our machine, 
but we only use 28 threads without hyperthreading. 
For the largest graph, \texttt{Gsh}, we used a 2 socket machine with Intel's second 
generation Xeon scalable processor with 2.2 Ghz and 48 cores, equipped with 6TB of 
Intel Optane PMM~\cite{optane} (byte-addressable memory technology).
Our GPU platforms are NVIDIA GTX 1080Ti (11GB memory) and Tesla
V100 (32GB memory) GPUs with CUDA 9.0. Unless specified
otherwise, GPU results reported are on V100.

RStream writes its intermediate data to the SSD, 
whereas other frameworks run all applications in memory.
We exclude preprocessing time and only report the computation
time (on the CPU or GPU) as an average of 3 runs.
\hlc{We also exclude the time to transfer data from CPU to GPU 
as it is trivial compared to the GPU compute time.}

\begin{table*}
\captionsetup[subtable]{aboveskip=3pt,belowskip=3pt}
	\centering
	\label{tab:combine}

	\begin{subtable}{.355\linewidth}
	\centering
	\resizebox{\textwidth}{!}{
	\begin{tabular}{c|rrrrr}
		\Xhline{2\arrayrulewidth}
		\bf{Input}   & \bf{G-Miner} & \bf{GAP} & \bf{PA-CPU} & \bf{DistTC-GPU} & \bf{PA-GPU} \\ \hline
		\texttt{Lj}  & 5.2     & 0.5      & 0.6         & 0.07                & \bf{0.06}   \\
		\texttt{Or}  & 13.3    & 4.2      & 3.9         & 0.3                 & \bf{0.2}    \\
		\texttt{Tw}  & 1067.7  & 40.1     & 38.8        & 4.3                 & \bf{2.9}    \\
		\Xhline{2\arrayrulewidth}
	\end{tabular}}
	\caption{TC. \hlc{GM: G-Miner.}}
	\label{tab:tc}
	\end{subtable}
\quad
	\begin{subtable}{.265\linewidth}
	\resizebox{0.92\textwidth}{!}{
	\begin{tabular}{c|rrrr}
		\Xhline{2\arrayrulewidth}
		\bf{Input}  & \bf{KClist} & \bf{PA-CPU} & \bf{PA-GPU} \\ \hline
		\texttt{Lj} & \bf{1.9}    & 26.3        & 2.3         \\
		\texttt{Or} & \bf{4.1}    & 82.3        & 4.3         \\
		\texttt{Tw} & \bf{628}    & 28165       & 1509        \\
		\Xhline{2\arrayrulewidth}
	\end{tabular}}
	\caption{4-CF.}
	\label{tab:4-clique}
	\end{subtable}
\quad
	\begin{subtable}{.33\linewidth}
	\resizebox{\textwidth}{!}{
	\begin{tabular}{c|rrrr}
		\Xhline{2\arrayrulewidth}
		\bf{Input}  & \bf{PGD} & \bf{PA-CPU} & \bf{PGD-GPU$^\dagger$} & \bf{PA-GPU} \\ \hline
		\texttt{Lj} & 12.7     & 19.5        & $\sim$\bf{1.4}            & 1.7    \\
		\texttt{Or} & 46.9     & 175         & $\sim$\bf{7.7}            & 18.0   \\
		\texttt{Tw}	& 1883     & 9388        & -                         & 1163   \\
		\Xhline{2\arrayrulewidth}
	\end{tabular}}
	\caption{3-MC.}
	\label{tab:3-motif}
	\end{subtable}

	\begin{subtable}{0.7\linewidth}
	\centering
	\resizebox{\textwidth}{!}{%
	\begin{tabular}{c|rrr|rrr|rrr|rrr}
		\Xhline{2\arrayrulewidth}
		\multicolumn{1}{l|}{} & \multicolumn{3}{c|}{\bf{Mico}} & \multicolumn{3}{c|}{\bf{Patent}} & \multicolumn{3}{c|}{\bf{Youtube}} & \multicolumn{3}{c}{\bf{PDB}} \\ \hline
		$\sigma$ & \bf{DG} & \bf{PA-CPU} & \bf{PA-GPU} & \bf{DG} & \bf{PA-CPU} & \bf{PA-GPU} & \bf{DG} & \bf{PA-CPU} & \bf{PA-GPU} & \bf{DG} & \bf{PA-CPU} & \bf{PA-GPU} \\ \hline
		300      & 52.2    & 3.9         & \bf{0.6}    & 19.9    & 14.7        & \bf{2.7}    & -       & \bf{96.9}   & -           & 281.4   & \bf{63.7} & -             \\
		500      & 52.9    & 3.6         & \bf{0.5}    & 18.7    & 15.8        & \bf{2.7}    & -       & \bf{97.7}   & -           & 279.5   & \bf{65.6} & -             \\
		1000     & 59.1    & 3.0         & \bf{0.4}    & 18.6    & 18.1        & \bf{2.7}    & -       & \bf{98.0}   & -           & 274.5   & \bf{73.4} & -             \\
		5000     & 58.1    & 2.4         & \bf{0.2}    & 18.4    & 27.0        & \bf{1.7}    & -       & \bf{102.3}  & -           & 322.9   & \bf{145.3}& -             \\
		\Xhline{2\arrayrulewidth}
	\end{tabular}}
	\caption{3-FSM. DG: DistGraph.}
	\label{tab:3-fsm}
	\end{subtable}
\quad
	\begin{subtable}{0.26\linewidth}
	\centering
	\resizebox{0.66\textwidth}{!}{%
	\begin{tabular}{c|rrr}
		\Xhline{2\arrayrulewidth}
		$\sigma$ & \bf{DG} & \bf{PA-CPU} \\ \hline
		15K   & \bf{129.0} & 438.9\\
		20K   & \bf{81.9}  & 224.7\\
		30K   & \bf{26.2}  & 31.9 \\ 
		\Xhline{2\arrayrulewidth}
	\end{tabular}}
	\caption{4-FSM for \texttt{Patent}.}
	\label{tab:4-fsm}
	\end{subtable}
	\vspace{-0.1cm}
	\caption{\small Execution time (sec) of Pangolin (PA) and hand-optimized solvers ($\sigma$: minimum support).
	PA-GPU and DistTC-GPU are on V100 GPU; PGD-GPU is on Titan Black GPU; rest are on 28 core CPU.
	$^\dagger$PGD-GPU results are reported from their paper.}
	\vspace{-0.3cm}
\end{table*}

\begin{table}
\captionsetup[subtable]{aboveskip=3pt,belowskip=3pt}
	\begin{subtable}{0.5\linewidth}
	\centering
	\resizebox{\textwidth}{!}{
	\begin{tabular}{c|rrr}
		\Xhline{2\arrayrulewidth}
		     & \bf{AM$^\dagger$} & \bf{PA-CPU} & \bf{PA-GPU} \\ \hline
		TC   & 0.04  & 0.02  & \bf{0.001}\\
		3-MC & 0.12  & 0.20  & \bf{0.02} \\
		4-MC & 22.0  & 175.6 & \bf{5.3}  \\ 
		5-CF & 11.4  & 60.5  & \bf{9.7}  \\ 
		\Xhline{2\arrayrulewidth}
	\end{tabular}}
	\caption{\texttt{Mi}.}
	\label{tab:mi}
	\end{subtable}
\quad
	\begin{subtable}{0.41\linewidth}
	\centering
	\resizebox{0.87\textwidth}{!}{
	\begin{tabular}{c|rr}
		\Xhline{2\arrayrulewidth}
		      & \bf{AM$^\dagger$} & \bf{PA}  \\ \hline
		TC    & 4966  & 139.3 \\
		3-CF  & -     & 659.3 \\
		4-CF  & 45399 & 23475 \\
		\Xhline{2\arrayrulewidth}
	\end{tabular}}
	\caption{\texttt{Gsh}.}
	\label{tab:gsh}
	\end{subtable}

	\caption{\small Execution time (sec) of Pangolin (PA) and AutoMine (AM). 
	Pangolin for \texttt{Gsh} is evaluated on Intel Optane-PMM machine.
	$^\dagger$AutoMine results are reported from its paper.}
	\vspace{-0.18cm}
	\label{tab:automine}
\end{table}

\subsection{GPM Frameworks}\label{subsect:frameworks-compare}
\hlc{\cref{tab:compare} reports the execution time of 
Arabesque, RStream, Kaleido, Fractal, and Pangolin. 
The execution time of G-Miner and AutoMine is reported 
in \cref{tab:tc} and \cref{tab:automine} respectively
(because it does not have other applications or datasets respectively). 
Note that Kaleido and AutoMine results on 28-core and 20-core CPU, 
respectively, are reported from their papers. 
We evaluate the rest on our 28-core CPU, 
except that we evaluate Pangolin for {\tt gsh} on 48-core CPU. 
Fractal and AutoMine use DFS-based exploration, whereas 
the rest use BFS-based exploration.
Pangolin is an order-of-magnitude faster than Arabesque, RStream, Fractal, and G-Miner.
Pangolin outperforms Kaleido in all cases except 4-MC on \texttt{patent}.
Pangolin on CPU is comparable or slower than AutoMine 
but outperforms it by exploiting the GPU.}

\hlc{For small inputs (e.g., TC and $3$-CF with \texttt{Mi}),
Arabesque suffers non-trivial overhead due to the startup cost of Giraph.
For large graphs, however, due to lack of algorithmic 
(e.g., eager pruning and customized pattern classification)
and data structure optimizations, it is also slower than Pangolin.}
On average, Pangolin is 49$\times$ faster than Arabesque.

For RStream, the number of partitions $P$ is a key performance knob.
For each configuration, we choose $P$ to be the 
best performing one among 10, 20, 50, and 100.
RStream only supports edge-induced exploration and
does not support pattern-specific optimization.
This results in extremely large search spaces for CF 
and MC because there are many more edges than vertices.
In addition, RStream does not scale well because of the 
intensive use of \texttt{mutex} locks for updating shared data.
Lastly, Pangolin avoids inefficient data structures and
expensive redundant computation (isomorphism test) used by RStream.
Pangolin is 88$\times$ faster than RStream on average
(Kaleido~\cite{Kaleido} also observes that RStream is slower than Arabesque).

On average, Pangolin is 2.6$\times$ faster than Kaleido
(7.4$\times$, 3.3$\times$, 2.4$\times$, and 1.6$\times$
for TC, CF, MC, and FSM respectively).
This is mainly due to DAG construction and customized pattern
classification in Pangolin.

\hlc{
Pangolin is on average 80$\times$ faster than Fractal.
Fractal is built on Spark and suffers from overheads due to it.
More importantly, some optimizations in hand-optimized DFS-based 
applications like PGD~\cite{PGD} and KClist~\cite{KClique} 
are not supported in Fractal, which limits its performance.
}

\hlc{
AutoMine uses a key optimization~\cite{PGD,KClique}
to remove redundant computation that 
can only be enabled in DFS-based exploration.
Due to this, 
when pattern size $k$ is large like in 5-CF and 4-MC,
AutoMine is faster than Pangolin. 
However, since Pangolin uses BFS-based
exploration which easily enables GPU acceleration, 
Pangolin on GPU is on average 5.8$\times$ faster than AutoMine.
It is not clear how to enable DFS mode for GPU efficiently, 
especially when $k$ is large.
Note that for all the applications, AutoMine can only do counting 
but not listing, because it has no automorphism test during extension
(instead it uses post-processing to address the multiplexity issue).
FSM in AutoMine uses frequency (which is not anti-monotonic) instead of domain support,
and thus it is not comparable to FSM in Pangolin.}

\subsection{Hand-Optimized GPM Applications}\label{sect:hand-compare}

We compare hand-optimized implementations with Pangolin on CPU and GPU.
We report
results for the largest datasets supported on our platform 
for each application.
\hlc{Note that all hand-optimized applications involve 
substantially more programming effort than Pangolin ones.
Hand-optimized TC has 4$\times$ more lines of code (LoC) than 
Pangolin TC.
The other hand-optimized applications have one or two orders 
of magnitude more LoC than Pangolin ones.}

\hlc{In \cref{tab:tc}, we compare with GAP~\cite{GAPBS} 
and DistTC~\cite{DistTC}, the state-of-the-art TC implementations on 
CPU and GPU, respectively. 
It is clear from \cref{tab:compare} and \cref{tab:tc} that}
TC implementations in existing GPM frameworks are orders of 
magnitude slower than the hand-optimized implementation in GAP.
In contrast, \hlc{Pangolin performs similar to GAP 
on the same CPU. 
Pangolin is also faster than DistTC on the same GPU
due to its embedding list data structure, 
which has better load balance and memory access behavior.}

\cref{tab:4-clique} compares our $4$-clique with KClist~\cite{KClique}, 
\hlc{the state-of-the-art CF implementation}.
Pangolin is 10 to 20$\times$ slower than KClist on the CPU, although 
GPU acceleration of Pangolin significantly reduces the performance gap.
This is because KClist constructs a shrinking local graph for each edge, 
which significantly reduces the search space. 
This optimization can only enabled in the DFS exploration. 
In \cref{tab:3-motif}, we observe the same trend for 
3-MC compared with \hlc{PGD, 
the state-of-the-art MC solver for multicore CPU~\cite{PGD} and GPU~\cite{Rossi}.
Note that PGD can only do counting, but not listing, as it only counts
some of the patterns and the other patterns' counts are calculated
directly using some formulas. In contrast, MC in Pangolin can do
both counting and listing. Another limitation of PGD is that it 
can only handle 3-MC and 4-MC, while Pangolin handles arbitrary $k$.
As PGD for GPU (PGD-GPU)~\cite{Rossi} is not released,
we estimate PGD-GPU performance using their reported 
speedup~\cite{Rossi} on Titan Black GPU. 
Pangolin-GPU is 20\% to 130\% slower.}

\cref{tab:3-fsm} and \cref{tab:4-fsm} compares our 3-FSM and 4-FSM, 
respectively, with DistGraph~\cite{DistGraph,ParFSM}.
DistGraph supports both shared-memory and distributed platforms.
DistGraph supports a runtime parameter $\sigma$, which specifies
the minimum support, but we had to modify it to add the maximum size $k$.
On CPU, Pangolin outperforms DistGraph for 3-FSM in all cases, 
except for \texttt{Pa} with support 5K. 
For graphs that fit in the GPU memory (\texttt{Mi}, \texttt{Pa}), 
\hlc{Pangolin on GPU is 6.9$\times$ to 290$\times$ faster than DistGraph.
In comparison, the GPU implementation of DistGraph is only
4$\times$ to 9$\times$ faster than its CPU implementation~\cite{GpuFSM} 
(we are not able able to run their GPU code and we cannot compare with their reported results as they do not evaluate the same datasets).}
For 4-FSM, Pangolin is 22\% to 240\% slower than DistGraph. 
The slowdown is mainly due to the algorithmic differences:
DistGraph adopts DFS exploration and a recursive approach which reduces
computation and memory consumption, while Pangolin does BFS exploration.

\begin{figure}[t]
\centering
	\begin{minipage}[t]{0.48\linewidth}
	\includegraphics[width=\linewidth]{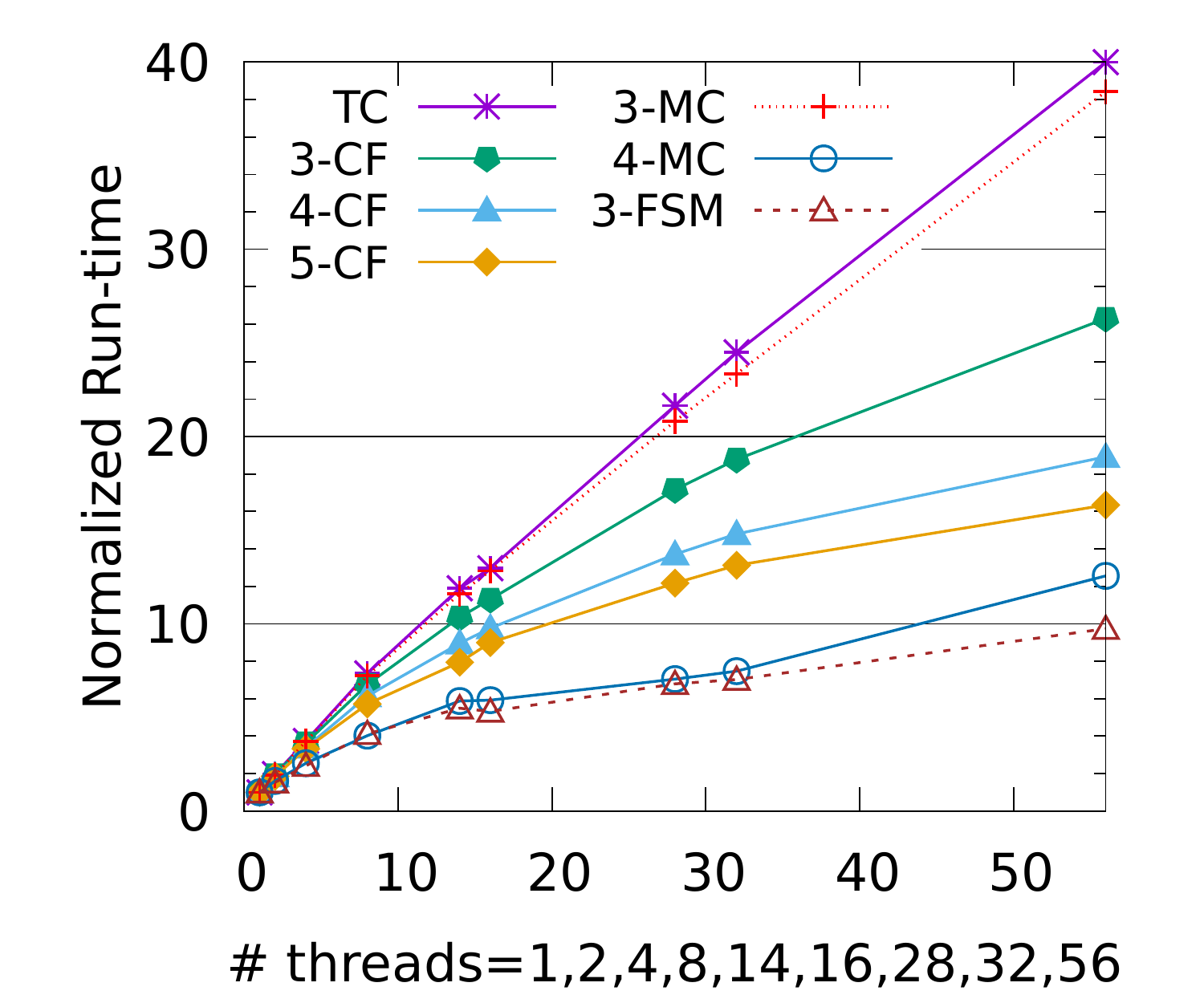}
	\caption{\scriptsize Strong scaling using \texttt{Yo} graph.
	$\sigma$=500 for FSM.}
	\label{fig:scale}
	\end{minipage}
\hfill
	\begin{minipage}[t]{0.48\linewidth}
	\includegraphics[width=\linewidth]{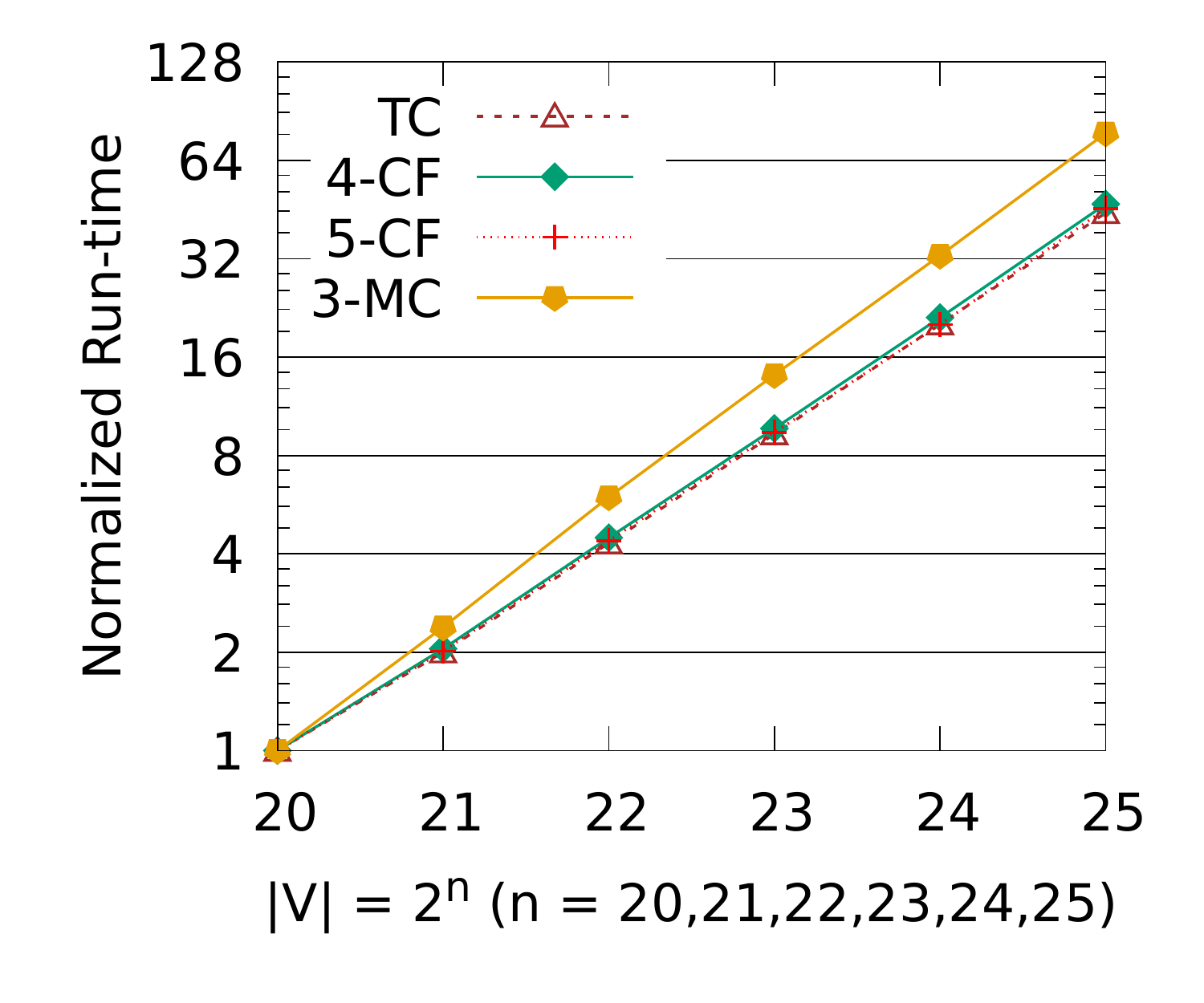}
	\caption{\scriptsize Execution time for RMAT graphs (log-log scale).}
	\label{fig:rmat}
	\end{minipage}
	\vspace{-0.15cm}
\end{figure}

\begin{figure*}[t]
	\begin{center}
		\includegraphics[width=0.92\textwidth]{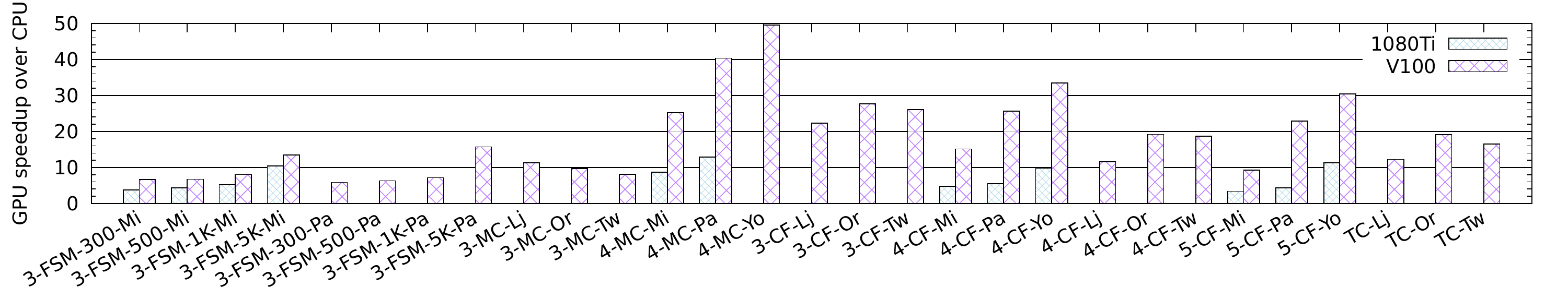}
		\vspace{-0.1cm}
		\caption{Speedup of Pangolin on GPU over Pangolin on 28-thread CPU. 
		Missing bars of 1080Ti are due to out of memory.}
		\vspace{-0.7cm}
		\label{fig:gpu}
	\end{center}
\end{figure*}

\subsection{Scalability and GPU Performance}
\label{subsect:scaling}

Although Pangolin is an in-memory processing system, 
Pangolin can scale to very large datasets by using large memory systems.
To demonstrate this, we evaluate Pangolin on the Intel Optane PMM system 
and mine a very large real-world web crawl, \texttt{Gsh}. 
As shown in \cref{tab:gsh}, TC and 3-CF only take 2 and 11 minutes, respectively.
4-CF is much more compute and memory intensive, so it takes $\sim6.5$ hours. 
To the best of our knowledge, 
this is the largest graph dataset for which 4-CF has been mined.

\cref{fig:scale} illustrates how the performance of Pangolin applications
scales as the number of threads increases for different applications on \texttt{Yo}.
Pangolin achieves good scalability by utilizing
efficient, concurrent, scalable data structures and allocators.
For TC, we observe near linear speedup over single-thread execution.
In contrast, FSM's scalability suffers 
due to the overheads of computing domain support.

\hlc{To test weak scaling, we use the RMAT graph generator~\cite{Khorasani}
to generate graphs with vertices $|V|$ from $2^{20}$ to $2^{25}$ 
and 
average degree $\overline{d}=20$.
\cref{fig:rmat} reports the execution time normalized to that of \texttt{rmat20}
(log-log scale). 
The execution time grows exponentially as the graph size increases
because the enumeration search space grows exponentially.}

\cref{fig:gpu} illustrates speedup of Pangolin
applications on GPU over 28 threads CPU.
Note that due to the limited memory size,
GPUs fail to run some applications and inputs.
On average, 1080Ti and V100 GPUs achieve a speedup
of 6$\times$ and 15$\times$ respectively over the CPU execution. 
Specifically, we observe substantial speedup on CF and MC. For example,
the V100 GPU achieves 50$\times$ speedup on 4-MC for \texttt{Yo},
demonstrating the suitability of GPUs for these compute intensive applications.


\begin{figure}[t]
	\begin{center}
		\includegraphics[width=0.46\textwidth]{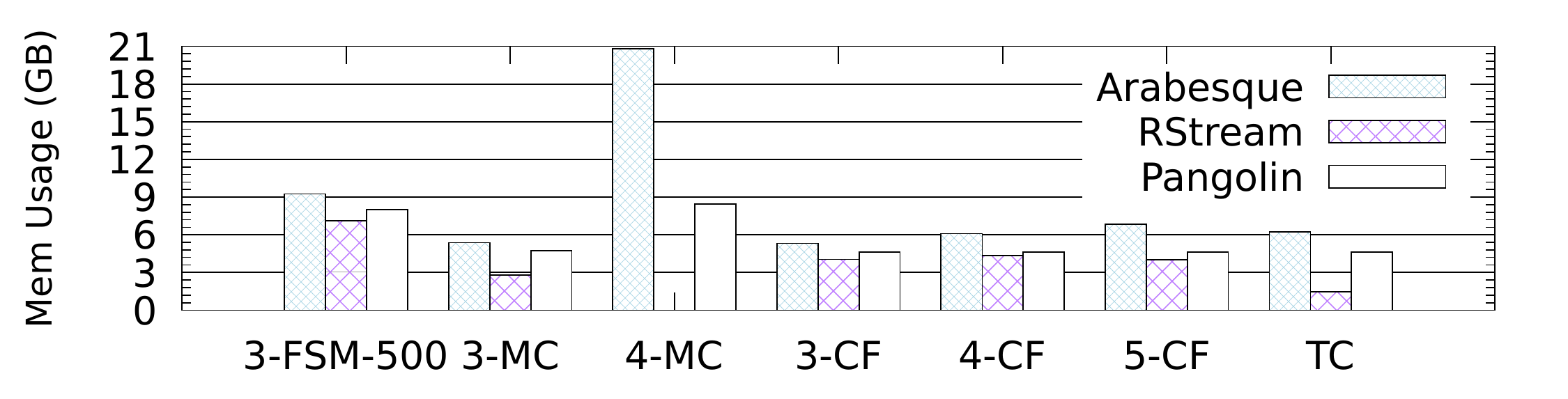}
		\vspace{0cm}
		\caption{\small Peak memory usage in Arabesque,
			RStream, and Pangolin for \texttt{Pa}
			(4-MC in RStream runs out of memory).}
		\vspace{-0.5cm}
		\label{fig:memory}
	\end{center}
\end{figure}

\subsection{Memory Consumption}\label{subsect:memory}
The peak memory consumption for Arabesque,
RStream, and Pangolin is illustrated in \cref{fig:memory}. 
\hlc{All systems are evaluated on the same 28-core CPU platform.}
We observe that Arabesque always requires
the most memory because it is implemented in Java using
Giraph~\cite{Giraph} framework that allocates a huge amount of memory.
In contrast, Pangolin avoids this overhead
and reduces memory usage. Since Pangolin does
in-memory computation, it is expected to consume much more memory
than RStream which stores its embeddings in disk. However, we find
that the difference in memory usage is trivial because aggressive search space
pruning and customized pattern classification significantly
reduce memory usage. Since this small memory cost brings substantial
performance improvement, we believe Pangolin makes a reasonable trade-off.
For 4-MC, RStream runs out of memory due to its edge-induced
exploration (Arabesque and Pangolin are using vertex-induced exploration).
\begin{figure}[t]
\centering
\captionsetup[subfigure]{aboveskip=-1pt,belowskip=0pt}
	\begin{subfigure}[t]{0.235\textwidth}
	\centering
		\includegraphics[width=\linewidth]{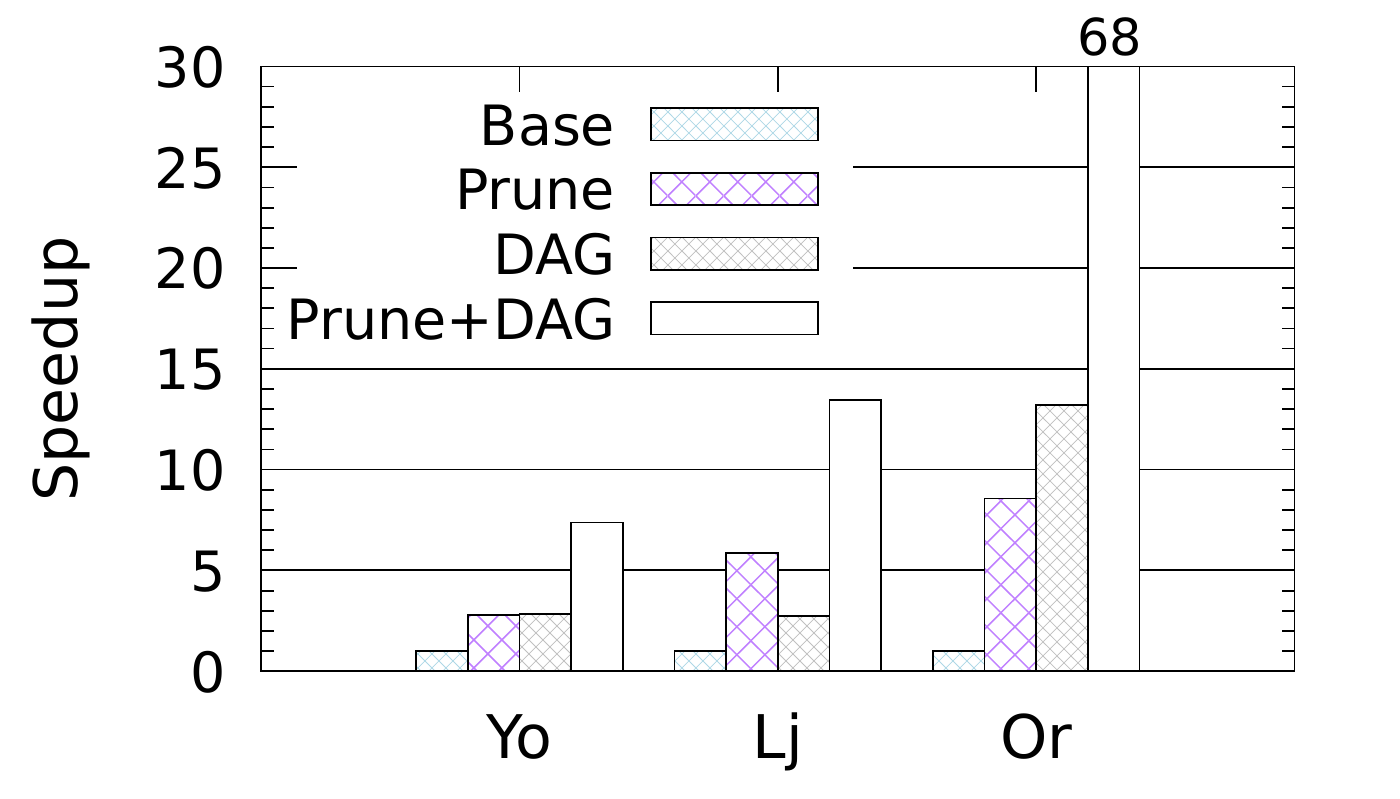}
		\caption{\scriptsize 4-CF {(pruning)}}
		\label{fig:prune}
	\end{subfigure}
	\begin{subfigure}[t]{0.234\textwidth}
	\centering
		\includegraphics[width=\linewidth]{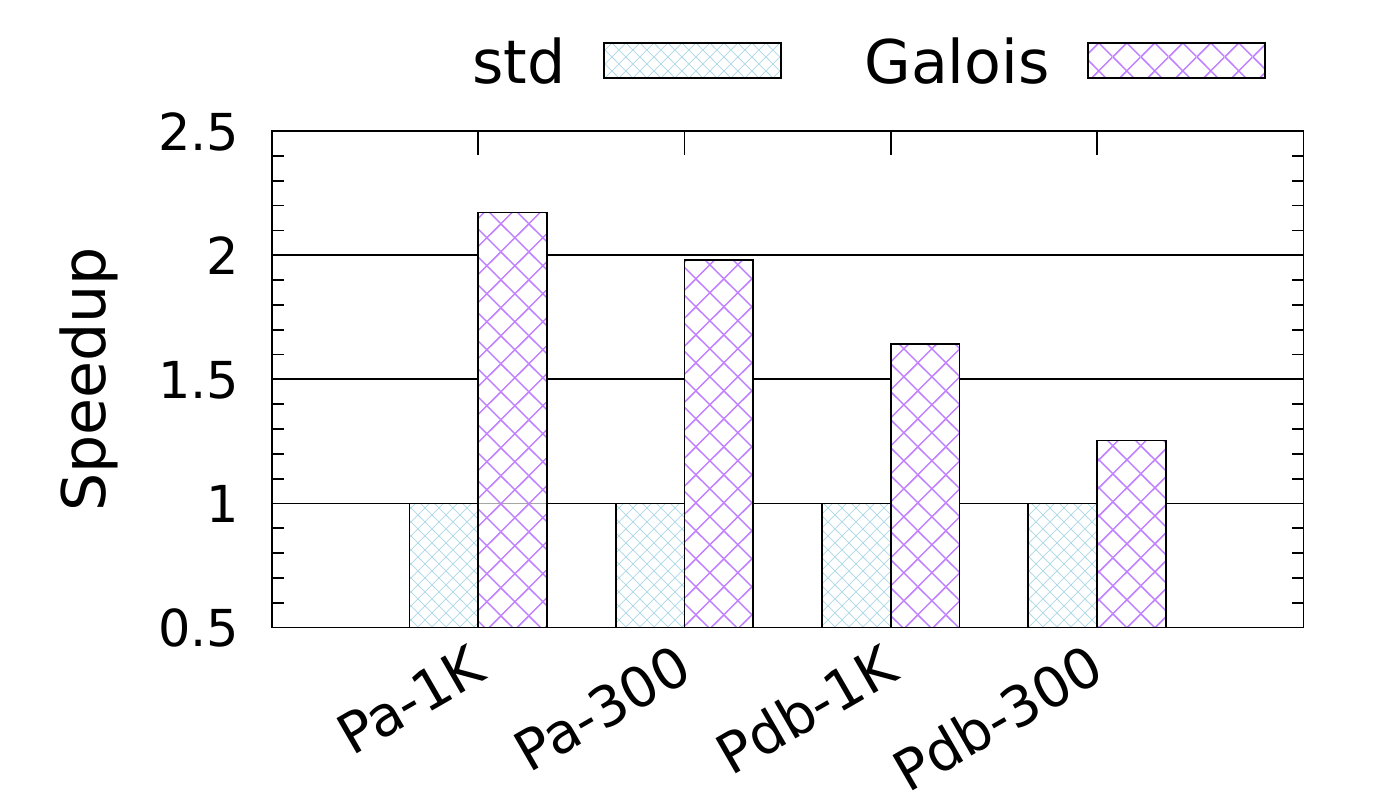}
		\caption{\scriptsize 3-FSM {(Galois allocator)}}
		\label{fig:alloc}
	\end{subfigure}
	
	\begin{subfigure}[t]{0.235\textwidth}
	\centering
		\includegraphics[width=\linewidth]{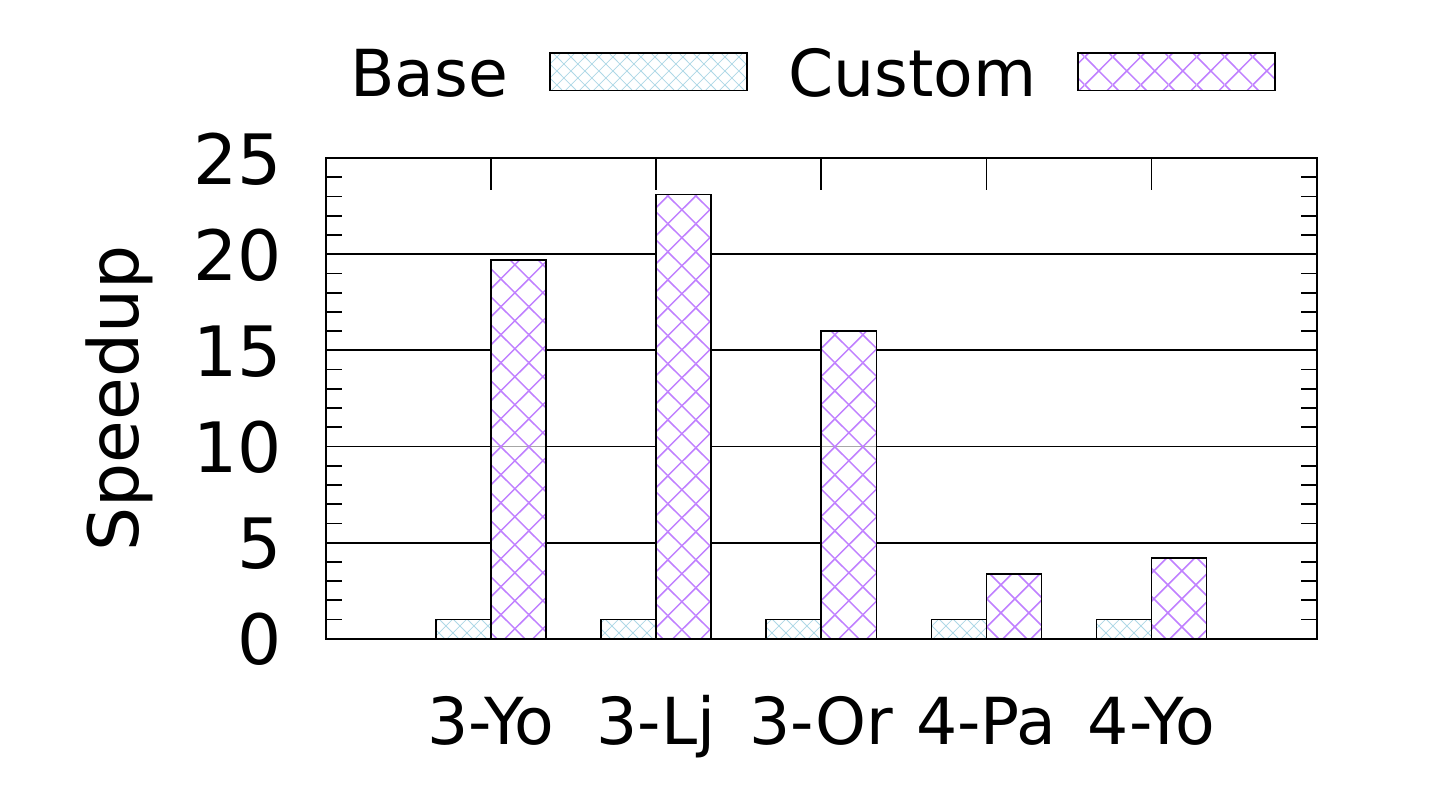}
		\caption{\scriptsize $k$-MC {(customized patterns)}}
		\label{fig:custom}
	\end{subfigure}
	\hfill
	\begin{subfigure}[t]{0.234\textwidth}
	\centering
		\includegraphics[width=\linewidth]{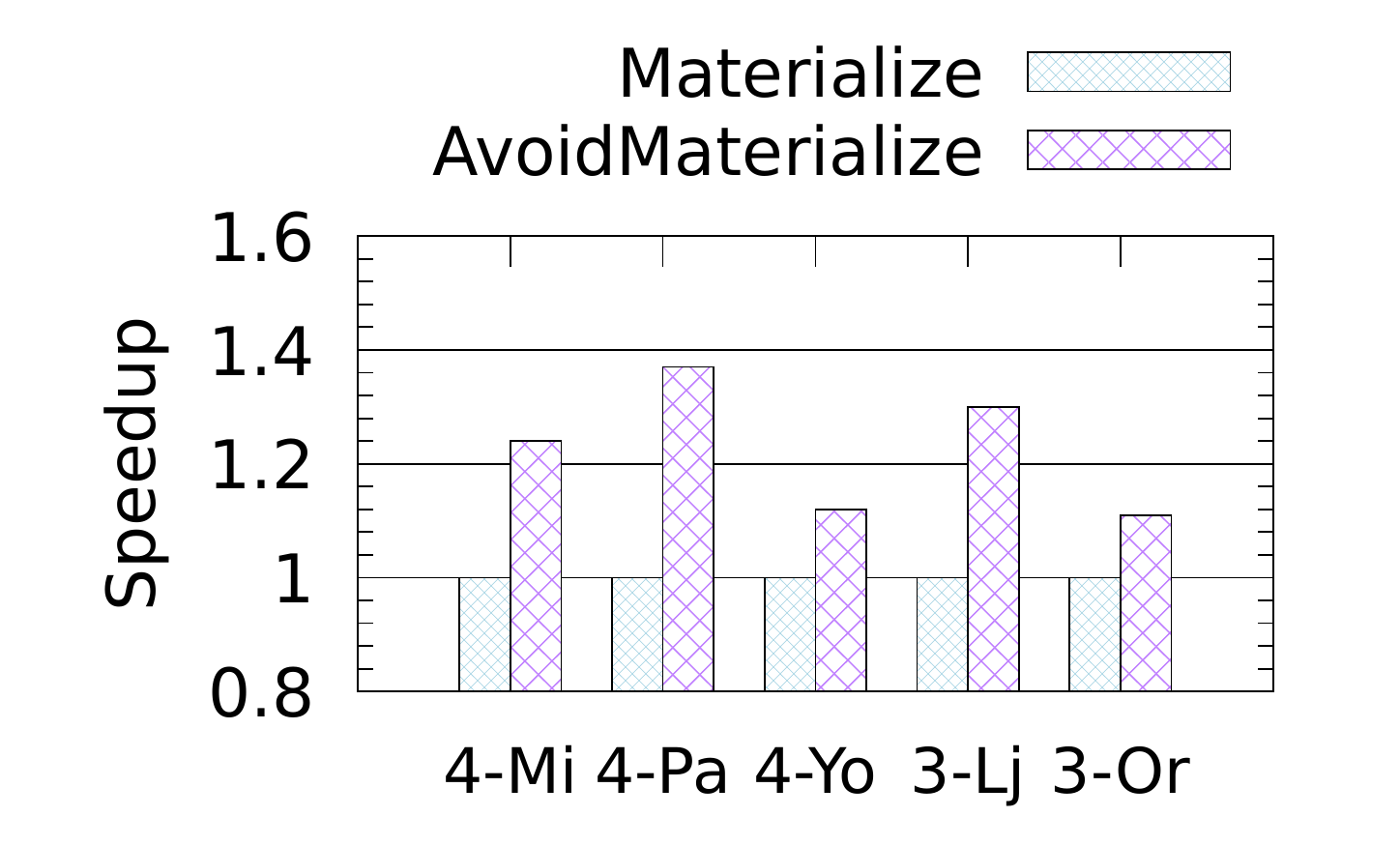}
		\caption{\scriptsize $k$-MC {(avoid materialization)}}
		\label{fig:eager}
	\end{subfigure}
\caption{\small Speedup due to various optimizations:
(\subref{fig:prune}) eager pruning and DAG;
(\subref{fig:alloc}) Galois scalable memory allocator;
(\subref{fig:custom}) customized pattern classification;
(\subref{fig:eager}) avoiding materialization.}
\vspace{-0.1cm}
\end{figure}

\begin{figure}[t]
\centering
\captionsetup[subfigure]{aboveskip=-1pt,belowskip=0pt}
\begin{subfigure}[t]{0.238\textwidth}
	\centering
	\includegraphics[width=\textwidth]{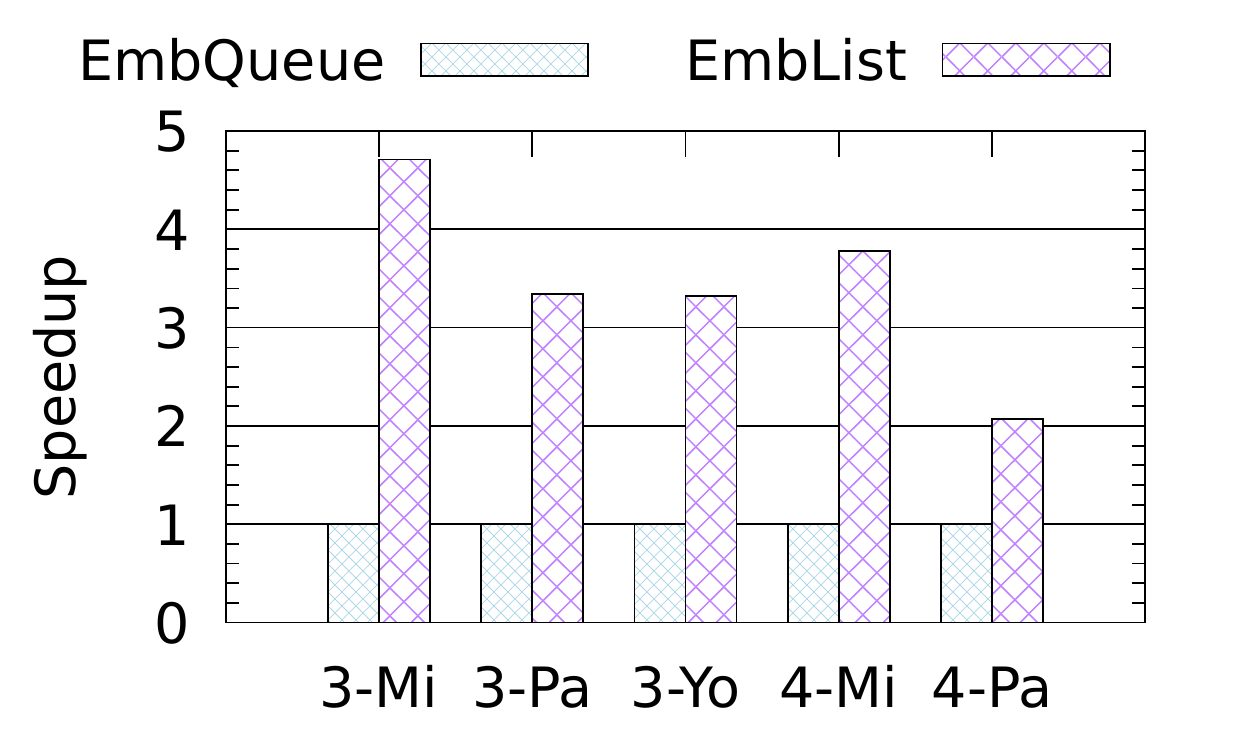}
	\caption{}
	\label{fig:list-vs-queue}
	\end{subfigure}
	\begin{subfigure}[t]{0.232\textwidth}
	\centering
	\includegraphics[width=\textwidth]{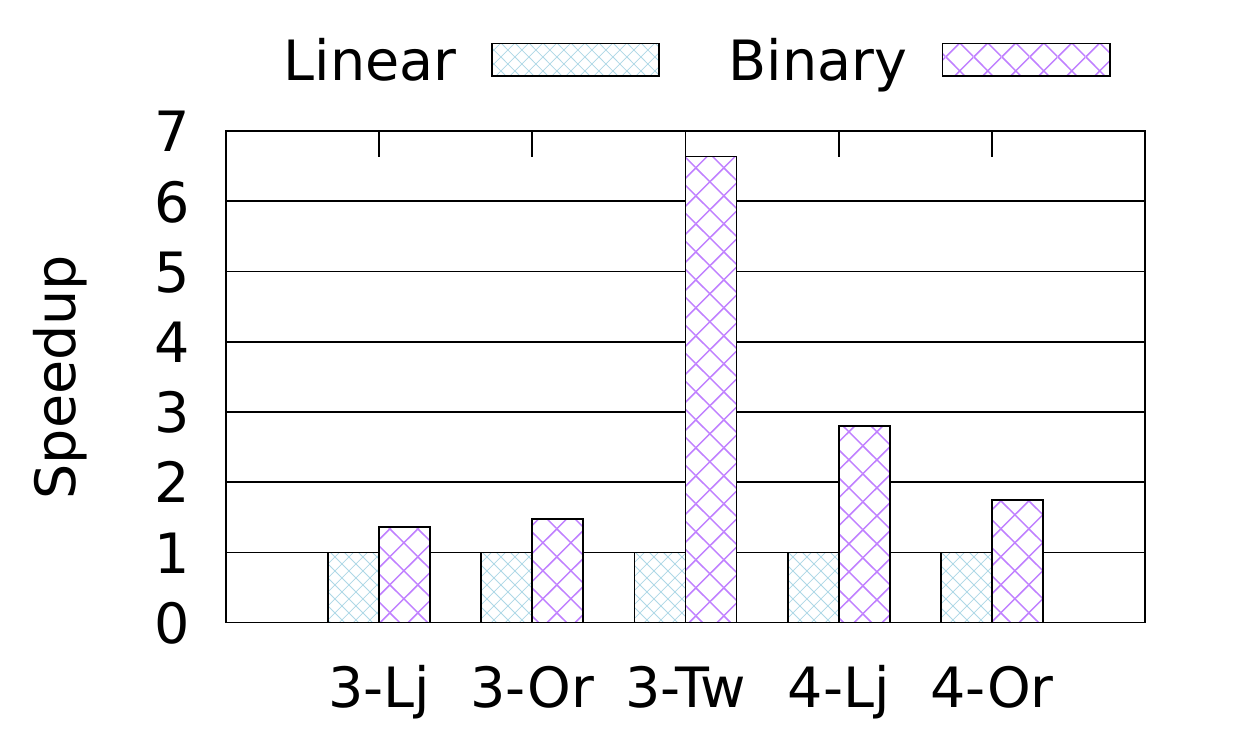}
	\caption{}
	\label{fig:search}
	\end{subfigure}
	\caption{\small $k$-MC speedup of (a) using embedding list (SoA+inspection-execution) 
		over using embedding queue (AoS) and (b) binary search over linear search.}
	\vspace{-0.36cm}
\end{figure}

\begin{figure}[t]
\begin{center}
	\includegraphics[width=0.4\textwidth]{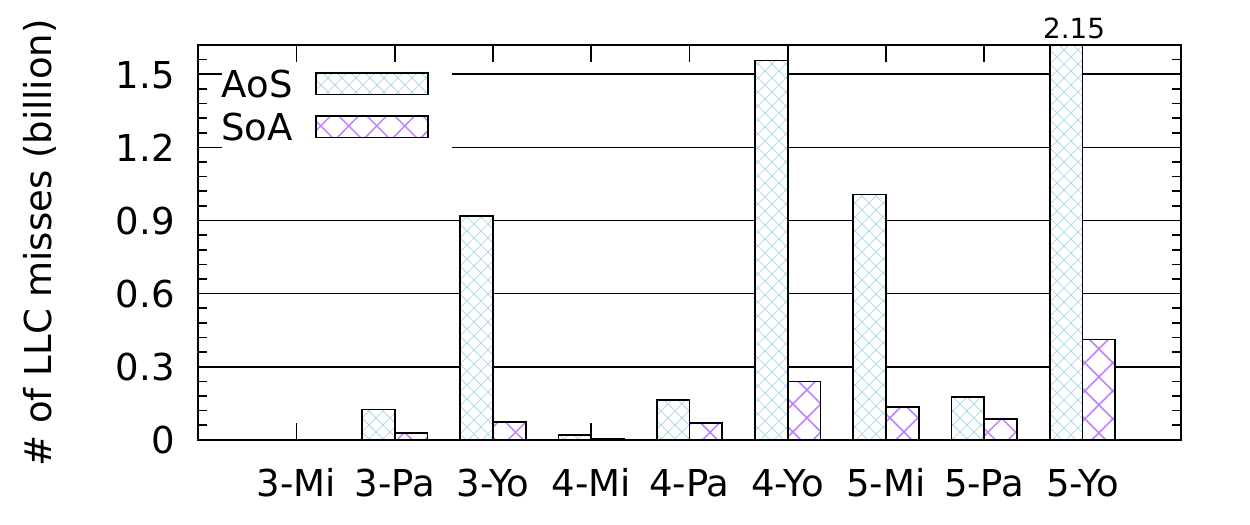}
	\vspace{-0cm}
	\caption{LLC miss counts in the vertex extension phase
	of $k$-CF using AoS and SoA for embeddings.}
	\vspace{-0.5cm}
	\label{fig:llc}
\end{center}
\end{figure}

\subsection{Impact of Optimizations}
We evaluate the performance improvement due to the optimizations
described in \cref{sect:app-opt} and \cref{sect:impl}. 
Due to lack of space, we present these comparisons 
only for the CPU implementations, 
but the results on the GPU are similar. 
\cref{fig:prune} shows the impact of orientation ({\em DAG}) 
and user-defined eager pruning ({\em Prune}) on 4-CF.
Both techniques significantly improve performance for TC (not shown) and CF.
\cref{fig:alloc} demonstrates the advantage of using Galois memory
allocators instead of \texttt{std} allocators. This is particularly important
for FSM as it requires intensive memory allocation for counting support.
\cref{fig:custom} illustrates that customized pattern classification
used in MC and FSM yields huge performance gains by eliding
expensive generic isomorphism tests. 
\cref{fig:eager} shows that materialization of temporary 
embeddings causes 11\% to 37\% slowdown for MC. 
This overhead exists in every application of Arabesque 
(and RStream), and is avoided in Pangolin.
In \cref{fig:list-vs-queue}, we evaluate the performance of 
our proposed embedding list data structure with SoA layout and
inspection-execution. Compared to the straight-forward embedding
queue (mimic the AoS implementation used in Arabesque and RStream),
the $k$-MC performance is 2.1$\times$ to 4.7$\times$ faster.
Another optimization is employing binary search for connectivity
check. \cref{fig:search} shows that binary search can achieve 
up to 6.6$\times$ speedup compared to linear search.
Finally, \cref{fig:llc} illustrates the last level cache (LLC) 
miss counts in the vertex extension phase of $k$-CF. We compare
two data structure schemes for the embeddings, AoS and SoA.
We observe a sharp reduction of LLC miss count by switching
from AoS to SoA. This further confirms that SoA has better
locality than AoS, due to the data reuse among embeddings.

\section{Related Work}\label{sect:relate}

\textbf{GPM Applications:}
Hand-optimized GPM applications target various platforms. For triangle counting,
Shun~\emph{et~al.}~\cite{Shun} present a parallel, cache-oblivious
TC solver on multicore CPUs that achieves good cache performance
without fine-tuning cache parameters. Load balancing is applied in
distributed TC solvers~\cite{Suri,PDTL} to evenly distribute workloads.
TriCore~\cite{TriCore} is a multi-GPU TC solver that uses binary search
to increase coalesced memory accesses, and it employs dynamic load balancing.

Chiba and Nishizeki (C\&N)~\cite{Arboricity} proposed an efficient 
$k$-clique listing algorithm which computes the subgraph induced 
by neighbors of each vertex, and then recurses on the subgraph. 
Danisch~\emph{et~al.}~\cite{KClique} refine the 
C\&N algorithm for parallelism and construct DAG 
using a core value based ordering to further reduce the search space.
PGD~\cite{PGD} counts 3 and 
4-motifs by leveraging a number of proven combinatorial arguments 
for different patterns. Some patterns (\emph{e.g.,} cliques) are counted 
first, and the frequencies of other patterns are obtained in constant 
time using these combinatorial arguments. Escape~\cite{ESCAPE} extends this 
approach to 5-vertex subgraphs and leverages DAG to reduce search space.

Frequent subgraph mining (FSM)~\cite{Huan} 
is one of the most important GPM applications.
gSpan~\cite{gSpan} is an efficient sequential FSM solver which implements a 
depth-first search (DFS) based on a lexicographic order called minimum DFS Code.
GraMi~\cite{GraMi} proposes an approach that finds only the minimal set of 
instances to satisfy the support threshold and avoids enumerating all instances. 
This idea has been adopted by most other frameworks. DistGraph~\cite{DistGraph} 
parallelizes gSpan for both shared-memory and distributed CPUs. Each worker thread
does the DFS walk concurrently. To balance workload, it introduces a customized
dynamic load balancing strategy which splits tasks on the fly and recomputes
the embedding list from scratch after the task is sent to a new worker.
Scalemine~\cite{Scalemine} solves FSM with a two-phase
approach, which approximates frequent subgraphs in phase-1, 
and uses collected information to compute the exact solution in phase-2.

Other important GPM applications includes \textit{maximal cliques}~\cite{MaximalClique},
\textit{maximum clique}~\cite{MaximumClique,Aboulnaga},
and \textit{subgraph listing}~\cite{SubgraphListing,CECI,DUALSIM,PathSampling,TurboFlux,Ma,Lai}.
They employ various optimizations to reduce 
computation and improve hardware efficiency.
They inspired our work to design a flexible interface for user-defined
optimizations. However, they achieve high performance at the cost of tremendous 
programming efforts, while Pangolin provides a unified model for ease of programming.

\textbf{GPM Frameworks:}
For ease-of-programming, GPM systems 
such as Arabesque~\cite{Arabesque}, RStream~\cite{RStream}, 
\hlc{G-Miner~\cite{G-Miner}}, and Kaleido~\cite{Kaleido} have been proposed. 
They provide a unified programming interface to the user
which simplifies application development. 
However, their interface is not flexible enough to 
enable application specific optimizations.
Instead of the BFS exploration used in these frameworks, 
Fractal~\cite{Fractal} employs a DFS strategy to enumerate subgraphs, 
which reduces memory footprint.
\hlc{AutoMine~\cite{AutoMine} is a compiler-based GPM system using DFS exploration.
In contrast, Pangolin uses the BFS approach that is inherently more load-balanced,
and is better suited for GPU acceleration.}
In the future, we plan to also support DFS exploration.
\hlc{EvoGraph~\cite{EvoGraph} is a GPU framework supporting both graph analytics 
and mining. However, as it is not designed specifically for GPM, 
many features such as automorphism and isomorphism test are  
not supported, which places a lot of burden on the programmer for complex GPM problems.}

\textbf{Approximate GPM:} 
There are approximate solvers for TC~\cite{DOULION,Rahman,Tsourakakis}, 
CF~\cite{Mitzenmacher,Jain}, MC~\cite{Slota,Bressan1}, and FSM~\cite{Approx}.
ASAP~\cite{ASAP} is an approximate GPM framework that supports various
GPM applications. It extends graph approximation theory to general 
patterns and incurs less than 5\% error. Since approximation reduces computation,
ASAP is much faster than exact frameworks like Arabesque, and scales to large graphs.
\hlc{Chen and Lui~\cite{MineOSN} propose another approximate GPM framework based on random walk.}
Compared to approximate solutions, Pangolin focuses on exact GPM and
achieves high performance without sacrificing accuracy.

\section{Conclusion}\label{sect:concl}
We present Pangolin, a high-performance, flexible
GPM system on shared-memory CPUs and GPUs.
Pangolin provides a simple
programming interface that enables the user to specify
eager enumeration search space pruning and
customized pattern classifications.
To exploit locality,
Pangolin uses an efficient structure of arrays (SoA)
for storing embeddings.
It avoids materialization of temporary embeddings
and blocks the schedule of
embedding exploration to reduce the memory usage.
It also uses inspection-execution and scalable memory allocators
to mitigate the overheads of dynamic memory allocation.
These application-specific and architectural optimizations
enable Pangolin to outperform \hlc{prior GPM frameworks,
Arabesque, RStream, and Fractal, by 49$\times$, 88$\times$, and 80$\times$}, on average, 
respectively, on the same 28-core CPU. Moreover, Pangolin on 
V100 GPU is 15$\times$ faster than that on the CPU on average.
Thus, Pangolin provides performance competitive with hand-optimized
implementations but with much better programming experience.
To mine 4-cliques in a web-crawl (gsh)
with 988 million vertices and 51 billion edges,
Pangolin takes $\sim6.5$ hours on a 48-core 
Intel Optane machine with 6 TB memory.

\bibliographystyle{abbrv}
\bibliography{references}

\begin{thebibliography}{10}

\bibitem{Scalemine}
E.~Abdelhamid, I.~Abdelaziz, P.~Kalnis, Z.~Khayyat, and F.~Jamour.
\newblock Scalemine: Scalable parallel frequent subgraph mining in a single
  large graph.
\newblock In {\em Proceedings of the International Conference for High
  Performance Computing, Networking, Storage and Analysis}, SC '16, pages
  61:1--61:12, Piscataway, NJ, USA, 2016. IEEE Press.

\bibitem{Aboulnaga}
A.~Aboulnaga, J.~Xiang, and C.~Guo.
\newblock Scalable maximum clique computation using mapreduce.
\newblock In {\em Proceedings of the 2013 IEEE International Conference on Data
  Engineering (ICDE 2013)}, ICDE '13, pages 74--85, Washington, DC, USA, 2013.
  IEEE Computer Society.

\bibitem{PGD}
N.~K. Ahmed, J.~Neville, R.~A. Rossi, and N.~Duffield.
\newblock Efficient graphlet counting for large networks.
\newblock In {\em ICDM}, pages 1--10, 2015.

\bibitem{Motifs1}
N.~Alon, P.~Dao, I.~Hajirasouliha, F.~Hormozdiari, and S.~Sahinalp.
\newblock Biomolecular network motif counting and discovery by color coding.
\newblock {\em Bioinformatics}, 24(13):241--249, 2008.

\bibitem{Alon}
N.~Alon, R.~Yuster, and U.~Zwick.
\newblock Color-coding: A new method for finding simple paths, cycles and other
  small subgraphs within large graphs.
\newblock In {\em Proceedings of the Twenty-sixth Annual ACM Symposium on
  Theory of Computing}, STOC '94, pages 326--335, New York, NY, USA, 1994. ACM.

\bibitem{Approx}
P.~Anchuri, M.~J. Zaki, O.~Barkol, S.~Golan, and M.~Shamy.
\newblock Approximate graph mining with label costs.
\newblock In {\em Proceedings of the 19th ACM SIGKDD International Conference
  on Knowledge Discovery and Data Mining}, KDD '13, pages 518--526, New York,
  NY, USA, 2013. ACM.

\bibitem{Complexity}
L.~{Babai}, W.~M. {Kantor}, and E.~M. {Luks}.
\newblock Computational complexity and the classification of finite simple
  groups.
\newblock In {\em 24th Annual Symposium on Foundations of Computer Science
  (sfcs 1983)}, pages 162--171, Nov 1983.

\bibitem{Memory}
A.~{Basak}, S.~{Li}, X.~{Hu}, S.~M. {Oh}, X.~{Xie}, L.~{Zhao}, X.~{Jiang}, and
  Y.~{Xie}.
\newblock Analysis and optimization of the memory hierarchy for graph
  processing workloads.
\newblock In {\em 2019 IEEE International Symposium on High Performance
  Computer Architecture (HPCA)}, pages 373--386, Feb 2019.

\bibitem{Locality}
S.~Beamer, K.~Asanovic, and D.~Patterson.
\newblock Locality exists in graph processing: Workload characterization on an
  ivy bridge server.
\newblock In {\em Proceedings of the 2015 IEEE International Symposium on
  Workload Characterization}, IISWC '15, pages 56--65, Washington, DC, USA,
  2015. IEEE Computer Society.

\bibitem{GAPBS}
S.~Beamer, K.~Asanovic, and D.~A. Patterson.
\newblock The {GAP} benchmark suite.
\newblock {\em CoRR}, abs/1508.03619, 2015.

\bibitem{Motif3}
A.~R. Benson, D.~F. Gleich, and J.~Leskovec.
\newblock Higher-order organization of complex networks.
\newblock {\em Science}, 353(6295):163--166, 2016.

\bibitem{Hoard}
E.~D. Berger, K.~S. McKinley, R.~D. Blumofe, and P.~R. Wilson.
\newblock Hoard: A scalable memory allocator for multithreaded applications.
\newblock In {\em Proceedings of the Ninth International Conference on
  Architectural Support for Programming Languages and Operating Systems},
  ASPLOS IX, pages 117--128, New York, NY, USA, 2000. ACM.

\bibitem{CECI}
B.~Bhattarai, H.~Liu, and H.~H. Huang.
\newblock Ceci: Compact embedding cluster index for scalable subgraph matching.
\newblock In {\em Proceedings of the 2019 International Conference on
  Management of Data}, SIGMOD '19, pages 1447--1462, New York, NY, USA, 2019.
  ACM.

\bibitem{gsh2015}
P.~Boldi and S.~Vigna.
\newblock The {W}eb{G}raph framework {I}: {C}ompression techniques.
\newblock In {\em Proc. of the Thirteenth International World Wide Web
  Conference (WWW 2004)}, pages 595--601, Manhattan, USA, 2004. ACM Press.

\bibitem{Bressan1}
M.~Bressan, F.~Chierichetti, R.~Kumar, S.~Leucci, and A.~Panconesi.
\newblock Counting graphlets: Space vs time.
\newblock In {\em Proceedings of the Tenth ACM International Conference on Web
  Search and Data Mining}, WSDM '17, pages 557--566, New York, NY, USA, 2017.
  ACM.

\bibitem{Bressan}
M.~Bressan, F.~Chierichetti, R.~Kumar, S.~Leucci, and A.~Panconesi.
\newblock Motif counting beyond five nodes.
\newblock {\em ACM Trans. Knowl. Discov. Data}, 12(4):48:1--48:25, Apr. 2018.

\bibitem{LonestarGPU}
M.~{Burtscher}, R.~{Nasre}, and K.~{Pingali}.
\newblock A quantitative study of irregular programs on gpus.
\newblock In {\em 2012 IEEE International Symposium on Workload
  Characterization (IISWC)}, pages 141--151, Nov 2012.

\bibitem{G-Miner}
H.~Chen, M.~Liu, Y.~Zhao, X.~Yan, D.~Yan, and J.~Cheng.
\newblock G-miner: An efficient task-oriented graph mining system.
\newblock In {\em Proceedings of the Thirteenth EuroSys Conference}, EuroSys
  ’18, New York, NY, USA, 2018. Association for Computing Machinery.

\bibitem{MineOSN}
X.~Chen and J.~C.~S. Lui.
\newblock Mining graphlet counts in online social networks.
\newblock {\em ACM Trans. Knowl. Discov. Data}, 12(4), Apr. 2018.

\bibitem{MaximalClique}
J.~Cheng, L.~Zhu, Y.~Ke, and S.~Chu.
\newblock Fast algorithms for maximal clique enumeration with limited memory.
\newblock In {\em Proceedings of the 18th ACM SIGKDD International Conference
  on Knowledge Discovery and Data Mining}, KDD '12, pages 1240--1248, New York,
  NY, USA, 2012. ACM.

\bibitem{Youtube}
X.~Cheng, C.~Dale, and J.~Liu.
\newblock Dataset for statistics and social network of youtube videos.
\newblock http://netsg.cs.sfu.ca/youtubedata/.

\bibitem{Falcon}
U.~Cheramangalath, R.~Nasre, and Y.~N. Srikant.
\newblock Falcon: A graph manipulation language for heterogeneous systems.
\newblock {\em ACM Trans. Archit. Code Optim.}, 12(4), Dec. 2015.

\bibitem{Arboricity}
N.~Chiba and T.~Nishizeki.
\newblock Arboricity and subgraph listing algorithms.
\newblock {\em SIAM J. Comput.}, 14(1):210--223, Feb. 1985.

\bibitem{Protein}
Y.-R. Cho and A.~Zhang.
\newblock Predicting protein function by frequent functional association
  pattern mining in protein interaction networks.
\newblock {\em Trans. Info. Tech. Biomed.}, 14(1):30--36, Jan. 2010.

\bibitem{KClique}
M.~Danisch, O.~Balalau, and M.~Sozio.
\newblock Listing k-cliques in sparse real-world graphs*.
\newblock In {\em Proceedings of the 2018 World Wide Web Conference}, WWW '18,
  pages 589--598, Republic and Canton of Geneva, Switzerland, 2018.
  International World Wide Web Conferences Steering Committee.

\bibitem{Gluon}
R.~Dathathri, G.~Gill, L.~Hoang, H.-V. Dang, A.~Brooks, N.~Dryden, M.~Snir, and
  K.~Pingali.
\newblock Gluon: A communication-optimizing substrate for distributed
  heterogeneous graph analytics.
\newblock In {\em Proceedings of the 39th ACM SIGPLAN Conference on Programming
  Language Design and Implementation}, PLDI 2018, pages 752--768, New York, NY,
  USA, 2018. ACM.

\bibitem{chemical}
M.~{Deshpande}, M.~{Kuramochi}, N.~{Wale}, and G.~{Karypis}.
\newblock Frequent substructure-based approaches for classifying chemical
  compounds.
\newblock {\em IEEE Transactions on Knowledge and Data Engineering},
  17(8):1036--1050, Aug 2005.

\bibitem{Fractal}
V.~Dias, C.~H.~C. Teixeira, D.~Guedes, W.~Meira, and S.~Parthasarathy.
\newblock Fractal: A general-purpose graph pattern mining system.
\newblock In {\em Proceedings of the 2019 International Conference on
  Management of Data}, SIGMOD '19, pages 1357--1374, New York, NY, USA, 2019.
  ACM.

\bibitem{BeyondTri}
E.~R. Elenberg, K.~Shanmugam, M.~Borokhovich, and A.~G. Dimakis.
\newblock Beyond triangles: A distributed framework for estimating 3-profiles
  of large graphs.
\newblock In {\em Proceedings of the 21th ACM SIGKDD International Conference
  on Knowledge Discovery and Data Mining}, KDD '15, pages 229--238, New York,
  NY, USA, 2015. ACM.

\bibitem{GraMi}
M.~Elseidy, E.~Abdelhamid, S.~Skiadopoulos, and P.~Kalnis.
\newblock Grami: Frequent subgraph and pattern mining in a single large graph.
\newblock {\em Proc. VLDB Endow.}, 7(7):517--528, Mar. 2014.

\bibitem{Analysis}
S.~Eyerman, W.~Heirman, K.~D. Bois, J.~B. Fryman, and I.~Hur.
\newblock Many-core graph workload analysis.
\newblock In {\em Proceedings of the International Conference for High
  Performance Computing, Networking, Storage, and Analysis}, SC '18, pages
  22:1--22:11, Piscataway, NJ, USA, 2018. IEEE Press.

\bibitem{GRAPE}
W.~Fan, J.~Xu, Y.~Wu, W.~Yu, J.~Jiang, Z.~Zheng, B.~Zhang, Y.~Cao, and C.~Tian.
\newblock Parallelizing sequential graph computations.
\newblock In {\em Proceedings of the 2017 ACM International Conference on
  Management of Data}, SIGMOD ’17, page 495–510, New York, NY, USA, 2017.
  Association for Computing Machinery.

\bibitem{Social}
K.~Faust.
\newblock A puzzle concerning triads in social networks: Graph constraints and
  the triad census.
\newblock {\em Social Networks}, 32(3):221 -- 233, 2010.

\bibitem{Garey}
M.~R. Garey and D.~S. Johnson.
\newblock {\em Computers and Intractability: A Guide to the Theory of
  NP-Completeness}.
\newblock W. H. Freeman \& Co., New York, NY, USA, 1979.

\bibitem{PDTL}
I.~{Giechaskiel}, G.~{Panagopoulos}, and E.~{Yoneki}.
\newblock {PDTL}: Parallel and distributed triangle listing for massive graphs.
\newblock In {\em 2015 44th International Conference on Parallel Processing},
  pages 370--379, Sep. 2015.

\bibitem{optane}
G.~Gill, R.~Dathathri, L.~Hoang, R.~Peri, and K.~Pingali.
\newblock Single machine graph analytics on massive datasets using intel optane
  {DC} persistent memory.
\newblock {\em CoRR}, abs/1904.07162, 2019.

\bibitem{Giraph}
Apache {G}iraph.
\newblock http://giraph.apache.org/, 2013.

\bibitem{PowerGraph}
J.~E. Gonzalez, Y.~Low, H.~Gu, D.~Bickson, and C.~Guestrin.
\newblock {PowerGraph: Distributed Graph-parallel Computation on Natural
  Graphs}.
\newblock In {\em Proceedings of the 10th USENIX Conference on Operating
  Systems Design and Implementation}, OSDI'12, pages 17--30, Berkeley, CA, USA,
  2012. USENIX Association.

\bibitem{Green}
O.~Green, P.~Yalamanchili, and L.-M. Mungu\'{\i}a.
\newblock Fast triangle counting on the gpu.
\newblock In {\em Proceedings of the 4th Workshop on Irregular Applications:
  Architectures and Algorithms}, IA3 '14, pages 1--8, Piscataway, NJ, USA,
  2014. IEEE Press.

\bibitem{Patent}
B.~H. Hall, J.~A. B., and T.~M.
\newblock The {NBER} patent citation data file: Lessons, insights and
  methodological tools.
\newblock http://www.nber.org/patents/, 2001.

\bibitem{DistTC}
L.~Hoang, V.~Jatala, X.~Chen, U.~Agarwal, R.~Dathathri, G.~Gill, and
  K.~Pingali.
\newblock {DistTC}: High performance distributed triangle counting.
\newblock In {\em HPEC 2019 23rd IEEE High Performance Extreme Computing, Graph
  Challenge}, September 2019.

\bibitem{MultiGraph}
C.~{Hong}, A.~{Sukumaran-Rajam}, J.~{Kim}, and P.~{Sadayappan}.
\newblock Multigraph: Efficient graph processing on gpus.
\newblock In {\em 2017 26th International Conference on Parallel Architectures
  and Compilation Techniques (PACT)}, pages 27--40, Sep. 2017.

\bibitem{TriCore}
Y.~{Hu}, H.~{Liu}, and H.~H. {Huang}.
\newblock Tricore: Parallel triangle counting on gpus.
\newblock In {\em SC18: International Conference for High Performance
  Computing, Networking, Storage and Analysis}, pages 171--182, Nov 2018.

\bibitem{Huan}
J.~{Huan}, W.~{Wang}, and J.~{Prins}.
\newblock Efficient mining of frequent subgraphs in the presence of
  isomorphism.
\newblock In {\em Third IEEE International Conference on Data Mining}, pages
  549--552, Nov 2003.

\bibitem{ASAP}
A.~P. Iyer, Z.~Liu, X.~Jin, S.~Venkataraman, V.~Braverman, and I.~Stoica.
\newblock Asap: Fast, approximate graph pattern mining at scale.
\newblock In {\em Proceedings of the 12th USENIX Conference on Operating
  Systems Design and Implementation}, OSDI'18, pages 745--761, Berkeley, CA,
  USA, 2018. USENIX Association.

\bibitem{Jain}
S.~Jain and C.~Seshadhri.
\newblock A fast and provable method for estimating clique counts using
  tur\'{a}n's theorem.
\newblock In {\em Proceedings of the 26th International Conference on World
  Wide Web}, WWW '17, pages 441--449, Republic and Canton of Geneva,
  Switzerland, 2017. International World Wide Web Conferences Steering
  Committee.

\bibitem{PathSampling}
M.~Jha, C.~Seshadhri, and A.~Pinar.
\newblock Path sampling: A fast and provable method for estimating 4-vertex
  subgraph counts.
\newblock In {\em Proceedings of the 24th International Conference on World
  Wide Web}, WWW '15, pages 495--505, Republic and Canton of Geneva,
  Switzerland, 2015. International World Wide Web Conferences Steering
  Committee.

\bibitem{Bliss}
T.~Junttila and P.~Kaski.
\newblock Engineering an efficient canonical labeling tool for large and sparse
  graphs.
\newblock In {\em Proceedings of the Meeting on Algorithm Engineering \&
  Expermiments}, pages 135--149, Philadelphia, PA, USA, 2007. Society for
  Industrial and Applied Mathematics.

\bibitem{GpuFSM}
R.~Kessl, N.~Talukder, P.~Anchuri, and M.~J. Zaki.
\newblock Parallel graph mining with gpus.
\newblock In {\em Proceedings of the 3rd International Conference on Big Data,
  Streams and Heterogeneous Source Mining: Algorithms, Systems, Programming
  Models and Applications - Volume 36}, BIGMINE'14, pages 1--16. JMLR.org,
  2014.

\bibitem{Khorasani}
F.~Khorasani, R.~Gupta, and L.~N. Bhuyan.
\newblock Scalable simd-efficient graph processing on gpus.
\newblock In {\em Proceedings of the 2015 International Conference on Parallel
  Architecture and Compilation (PACT)}, PACT ’15, page 39–50, USA, 2015.
  IEEE Computer Society.

\bibitem{DUALSIM}
H.~Kim, J.~Lee, S.~S. Bhowmick, W.-S. Han, J.~Lee, S.~Ko, and M.~H. Jarrah.
\newblock {DUALSIM}: Parallel subgraph enumeration in a massive graph on a
  single machine.
\newblock In {\em Proceedings of the 2016 International Conference on
  Management of Data}, SIGMOD '16, pages 1231--1245, New York, NY, USA, 2016.
  ACM.

\bibitem{TurboFlux}
K.~Kim, I.~Seo, W.-S. Han, J.-H. Lee, S.~Hong, H.~Chafi, H.~Shin, and G.~Jeong.
\newblock Turboflux: A fast continuous subgraph matching system for streaming
  graph data.
\newblock In {\em Proceedings of the 2018 International Conference on
  Management of Data}, SIGMOD '18, pages 411--426, New York, NY, USA, 2018.
  ACM.

\bibitem{Konect}
J.~Kunegis.
\newblock Konect: the koblenz network collection.
\newblock In {\em Proceedings of the 22nd International Conference on World
  Wide Web}, pages 1343--1350. ACM, 2013.

\bibitem{Lai}
L.~Lai, L.~Qin, X.~Lin, and L.~Chang.
\newblock Scalable subgraph enumeration in mapreduce.
\newblock {\em Proc. VLDB Endow.}, 8(10):974--985, June 2015.

\bibitem{SNAP}
J.~Leskovec.
\newblock Snap: Stanford network analysis platform, 2013.

\bibitem{MotifGPU}
W.~{Lin}, X.~{Xiao}, X.~{Xie}, and X.~{Li}.
\newblock Network motif discovery: A gpu approach.
\newblock In {\em 2015 IEEE 31st International Conference on Data Engineering},
  pages 831--842, April 2015.

\bibitem{SIMD-X}
H.~Liu and H.~H. Huang.
\newblock Simd-x: Programming and processing of graph algorithms on gpus.
\newblock In {\em Proceedings of the 2019 USENIX Conference on Usenix Annual
  Technical Conference}, USENIX ATC ’19, page 411–427, USA, 2019. USENIX
  Association.

\bibitem{CUDA-MEME}
Y.~Liu, B.~Schmidt, W.~Liu, and D.~L. Maskell.
\newblock Cuda-meme: Accelerating motif discovery in biological sequences using
  cuda-enabled graphics processing units.
\newblock {\em Pattern Recognition Letters}, 31(14):2170 -- 2177, 2010.

\bibitem{GraphLab}
Y.~Low, J.~Gonzalez, A.~Kyrola, D.~Bickson, C.~Guestrin, and J.~M. Hellerstein.
\newblock {{GraphLab}: A New Parallel Framework for Machine Learning}.
\newblock In {\em Proceedings Conf. Uncertainty in Artificial Intelligence},
  UAI '10, July 2010.

\bibitem{MaximumClique}
C.~Lu, J.~X. Yu, H.~Wei, and Y.~Zhang.
\newblock Finding the maximum clique in massive graphs.
\newblock {\em Proc. VLDB Endow.}, 10(11):1538--1549, Aug. 2017.

\bibitem{Ma}
S.~Ma, Y.~Cao, J.~Huai, and T.~Wo.
\newblock Distributed graph pattern matching.
\newblock In {\em Proceedings of the 21st International Conference on World
  Wide Web}, WWW '12, pages 949--958, New York, NY, USA, 2012. ACM.

\bibitem{Pregel}
G.~Malewicz, M.~H. Austern, A.~J. Bik, J.~C. Dehnert, I.~Horn, N.~Leiser, and
  G.~Czajkowski.
\newblock Pregel: A system for large-scale graph processing.
\newblock In {\em Proceedings of the ACM SIGMOD International Conference on
  Management of Data}, SIGMOD '10, pages 135--146, New York, NY, USA, 2010.
  ACM.

\bibitem{AutoMine}
D.~Mawhirter and B.~Wu.
\newblock Automine: Harmonizing high-level abstraction and high performance for
  graph mining.
\newblock In {\em Proceedings of the 27th ACM Symposium on Operating Systems
  Principles}, SOSP ’19, page 509–523, New York, NY, USA, 2019. Association
  for Computing Machinery.

\bibitem{Michael}
M.~M. Michael.
\newblock Scalable lock-free dynamic memory allocation.
\newblock In {\em Proceedings of the ACM SIGPLAN 2004 Conference on Programming
  Language Design and Implementation}, PLDI '04, pages 35--46, New York, NY,
  USA, 2004. ACM.

\bibitem{Motifs2}
R.~Milo, S.~Shen-Orr, S.~Itzkovitz, N.~Kashtan, D.~Chklovskii, and U.~Alon.
\newblock Network motifs: Simple building blocks of complex networks.
\newblock {\em Science}, 298(5594):824--827, 2002.

\bibitem{Mitzenmacher}
M.~Mitzenmacher, J.~Pachocki, R.~Peng, C.~Tsourakakis, and S.~C. Xu.
\newblock Scalable large near-clique detection in large-scale networks via
  sampling.
\newblock In {\em Proceedings of the 21th ACM SIGKDD International Conference
  on Knowledge Discovery and Data Mining}, KDD '15, pages 815--824, New York,
  NY, USA, 2015. ACM.

\bibitem{Galois}
D.~Nguyen, A.~Lenharth, and K.~Pingali.
\newblock A lightweight infrastructure for graph analytics.
\newblock In {\em Proceedings of the 24th ACM Symposium on Operating Systems
  Principles (SOSP)}, SOSP '13, pages 456--471, New York, NY, USA, 2013. ACM.

\bibitem{ESCAPE}
A.~Pinar, C.~Seshadhri, and V.~Vishal.
\newblock Escape: Efficiently counting all 5-vertex subgraphs.
\newblock In {\em Proceedings of the 26th International Conference on World
  Wide Web}, WWW '17, pages 1431--1440, Republic and Canton of Geneva,
  Switzerland, 2017. International World Wide Web Conferences Steering
  Committee.

\bibitem{Rahman}
M.~{Rahman} and M.~A. {Hasan}.
\newblock Approximate triangle counting algorithms on multi-cores.
\newblock In {\em 2013 IEEE International Conference on Big Data}, pages
  127--133, Oct 2013.

\bibitem{Rossi}
R.~A. Rossi and R.~Zhou.
\newblock Leveraging multiple gpus and cpus for graphlet counting in large
  networks.
\newblock In {\em Proceedings of the 25th ACM International on Conference on
  Information and Knowledge Management}, CIKM '16, pages 1783--1792, New York,
  NY, USA, 2016. ACM.

\bibitem{Schank}
T.~Schank and D.~Wagner.
\newblock Finding, counting and listing all triangles in large graphs, an
  experimental study.
\newblock In {\em Proceedings of the 4th International Conference on
  Experimental and Efficient Algorithms}, WEA'05, pages 606--609, Berlin,
  Heidelberg, 2005. Springer-Verlag.

\bibitem{Schneider}
S.~Schneider, C.~D. Antonopoulos, and D.~S. Nikolopoulos.
\newblock Scalable locality-conscious multithreaded memory allocation.
\newblock In {\em Proceedings of the 5th International Symposium on Memory
  Management}, ISMM '06, pages 84--94, New York, NY, USA, 2006. ACM.

\bibitem{EvoGraph}
D.~Sengupta and S.~L. Song.
\newblock Evograph: On-the-fly efficient mining of evolving graphs on gpu.
\newblock In J.~M. Kunkel, R.~Yokota, P.~Balaji, and D.~Keyes, editors, {\em
  High Performance Computing}, pages 97--119, Cham, 2017. Springer
  International Publishing.

\bibitem{SubgraphListing}
Y.~Shao, B.~Cui, L.~Chen, L.~Ma, J.~Yao, and N.~Xu.
\newblock Parallel subgraph listing in a large-scale graph.
\newblock In {\em Proceedings of the 2014 ACM SIGMOD International Conference
  on Management of Data}, SIGMOD '14, pages 625--636, New York, NY, USA, 2014.
  ACM.

\bibitem{Ligra}
J.~Shun and G.~E. Blelloch.
\newblock Ligra: A lightweight graph processing framework for shared memory.
\newblock In {\em Proceedings of the 18th ACM SIGPLAN Symposium on Principles
  and Practice of Parallel Programming (PPoPP)}, PPoPP '13, pages 135--146, New
  York, NY, USA, 2013. ACM.

\bibitem{Shun}
J.~{Shun} and K.~{Tangwongsan}.
\newblock Multicore triangle computations without tuning.
\newblock In {\em 2015 IEEE 31st International Conference on Data Engineering},
  pages 149--160, April 2015.

\bibitem{Slota}
G.~M. {Slota} and K.~{Madduri}.
\newblock Complex network analysis using parallel approximate motif counting.
\newblock In {\em 2014 IEEE 28th International Parallel and Distributed
  Processing Symposium}, pages 405--414, May 2014.

\bibitem{Suri}
S.~Suri and S.~Vassilvitskii.
\newblock Counting triangles and the curse of the last reducer.
\newblock In {\em Proceedings of the 20th International Conference on World
  Wide Web}, WWW '11, pages 607--614, New York, NY, USA, 2011. ACM.

\bibitem{DistGraph}
N.~Talukder and M.~J. Zaki.
\newblock A distributed approach for graph mining in massive networks.
\newblock {\em Data Min. Knowl. Discov.}, 30(5):1024--1052, Sept. 2016.

\bibitem{ParFSM}
N.~{Talukder} and M.~J. {Zaki}.
\newblock Parallel graph mining with dynamic load balancing.
\newblock In {\em 2016 IEEE International Conference on Big Data (Big Data)},
  pages 3352--3359, Dec 2016.

\bibitem{Arabesque}
C.~H.~C. Teixeira, A.~J. Fonseca, M.~Serafini, G.~Siganos, M.~J. Zaki, and
  A.~Aboulnaga.
\newblock Arabesque: A system for distributed graph mining.
\newblock In {\em Proceedings of the 25th Symposium on Operating Systems
  Principles}, SOSP '15, pages 425--440, New York, NY, USA, 2015. ACM.

\bibitem{Tsourakakis}
C.~E. {Tsourakakis}.
\newblock Fast counting of triangles in large real networks without counting:
  Algorithms and laws.
\newblock In {\em 2008 Eighth IEEE International Conference on Data Mining},
  pages 608--617, Dec 2008.

\bibitem{DOULION}
C.~E. Tsourakakis, U.~Kang, G.~L. Miller, and C.~Faloutsos.
\newblock Doulion: Counting triangles in massive graphs with a coin.
\newblock In {\em Proceedings of the 15th ACM SIGKDD International Conference
  on Knowledge Discovery and Data Mining}, KDD '09, pages 837--846, New York,
  NY, USA, 2009. ACM.

\bibitem{Voegele2017}
C.~Voegele, Y.-S. Lu, S.~Pai, and K.~Pingali.
\newblock Parallel triangle counting and k-truss identification using
  graph-centric methods.
\newblock In {\em IEEE/Amazon/DARPA GraphChallenge, IEEE HPEC,}, 2017.

\bibitem{RStream}
K.~Wang, Z.~Zuo, J.~Thorpe, T.~Q. Nguyen, and G.~H. Xu.
\newblock Rstream: Marrying relational algebra with streaming for efficient
  graph mining on a single machine.
\newblock In {\em Proceedings of the 12th USENIX Conference on Operating
  Systems Design and Implementation}, OSDI'18, pages 763--782, Berkeley, CA,
  USA, 2018. USENIX Association.

\bibitem{gSpan}
{Xifeng Yan} and {Jiawei Han}.
\newblock gspan: graph-based substructure pattern mining.
\newblock In {\em Proceedings of the 2002 IEEE International Conference on Data
  Mining}, pages 721--724, Dec 2002.

\bibitem{Kaleido}
C.~Zhao, Z.~Zhang, P.~Xu, T.~Zheng, and X.~Cheng.
\newblock Kaleido: An efficient out-of-core graph mining system on {A} single
  machine.
\newblock {\em CoRR}, abs/1905.09572, 2019.

\bibitem{Gemini}
X.~Zhu, W.~Chen, W.~Zheng, and X.~Ma.
\newblock Gemini: {A Computation-centric Distributed Graph Processing System}.
\newblock In {\em Proceedings of the 12th USENIX Conference on Operating
  Systems Design and Implementation}, OSDI'16, pages 301--316, Berkeley, CA,
  USA, 2016. USENIX Association.

\end{thebibliography}

\end{document}